%% file: LRstats_unequal_sens.tex
\documentclass[10pt]{iopart}
\usepackage{iopams}

\pdfoutput=1
\usepackage[square,sort&compress,numbers]{natbib}
\usepackage{graphicx}
\usepackage{hyperref}

\input{defs2.tex}
\newcommand{\dcc}{LIGO-P1400161}
\newcommand{\aei}{AEI-2014-039}

\makeatletter
\long\def\@caption#1[#2]#3{\par\begingroup
    \@parboxrestore
    \normalsize
    \@makecaption{\csname fnum@#1\endcsname}{\ignorespaces #3}\par
  \endgroup}
\long\def\@makecaption#1#2{\vskip \abovecaptionskip 
 \begin{lessindented}
 \item[]{\bf #1.} #2
 \end{lessindented}\vskip\belowcaptionskip}
\newenvironment{lessindented}{\begin{lessindented}}{\end{lessindented}}
\def\lessindented{\list{}{\itemsep=0\p@\labelsep=0\p@\itemindent=0\p@
   \labelwidth=0\p@\leftmargin=20\p@\topsep=0\p@\partopsep=0\p@
   \parsep=0\p@\listparindent=15\p@}\footnotesize\rm}

\makeatother

\begin{document}

\title[Line-robust CW statistics: safety in the case of unequal detector sensitivities]
{Line-robust statistics for continuous gravitational waves:
safety in the case of unequal detector sensitivities}

\author{David~Keitel, Reinhard~Prix}
\address{Albert-Einstein-Institut, Callinstrasse 38, 30167 Hannover, Germany}
\ead{\mailto{david.keitel@aei.mpg.de}, \mailto{reinhard.prix@aei.mpg.de}}

\vspace{10pt}
\begin{indented}
 \item[]published as \href{http://dx.doi.org/10.1088/0264-9381/32/3/035004}{Class. Quantum Grav. 32 035004}, 5 January 2015
 \item[]this version dated 15 January 2015
 \item[]LIGO document number: \dcc{}
\end{indented}

\begin{abstract}
The multi-detector $\F$-statistic is close to optimal for detecting continuous gravitational waves (CWs) in
Gaussian noise.
However, it is susceptible to false alarms from instrumental artefacts, for example quasi-monochromatic
disturbances ('lines'), which resemble a CW signal more than Gaussian noise.
In a recent paper~\cite{keitel2014:_linerobust}, a Bayesian model selection approach was used to derive
line-robust detection statistics for CW signals, generalising both the $\F$-statistic and the $\F$-statistic
consistency veto technique and yielding improved performance in line-affected data.
Here we investigate a generalisation of the assumptions made in that paper: if a CW analysis uses
data from two or more detectors with very different sensitivities, the line-robust statistics could be less
effective.
We investigate the boundaries within which they are still safe to use, in comparison with the $\F$-statistic.
Tests using synthetic draws show that the optimally-tuned version of the original line-robust statistic
remains safe in most cases of practical interest.
We also explore a simple idea on further improving the detection power and safety of these statistics, which
we however find to be of limited practical use.
\end{abstract}

\vspace{-1em}

\begin{indented}
 \item[] Keywords: gravitational waves, signal processing, Bayesian inference, neutron stars
\end{indented}

\vspace{-1em}


%
%
%
%

\vspace{6\baselineskip}

\section{Introduction}
\label{sec:introduction}

Continuous gravitational waves (CWs) are a potentially detectable class of astrophysical signals.
They are narrow-band in frequency and can typically be described by a relatively stable signal model over
years of observation~\cite{prix06:_cw_review,owen2009:_nsgw}.
In the frequency band covered by terrestrial interferometric detectors (such as LIGO~\cite{LIGORef:2009},
Virgo~\cite{VirgoRef:2011} and GEO\,600~\cite{GEORef:2010}) CWs would be produced by rotating neutron stars
with non-axisymmetric deformations~\cite{bildsten1998:_gwns, ushomirsky2000:_deformations, JMcD2013:_maxelastic}.

Most CW data-analysis methods assume a Gaussian distribution for the detector noise.
Indeed, in current interferometers this is a good description over most of the observation time and
frequency range.
(See, e.g., \cite{abbott2004:_geoligo,aasi13:_eathS5,behnke2013:_phdthesis}).
A standard detection statistic for CW signals in Gaussian noise is the \emph{$\F$-statistic}, corresponding to
a binary hypothesis test between a signal hypothesis and a Gaussian-noise hypothesis.
Originally derived as a maximum-likelihood detection statistic~\cite{jks98:_data,cutler05:_gen_fstat}, it was
also shown to follow from a Bayesian model-selection approach~\cite{prix09:_bstat}, using somewhat unphysical
priors for the signal-amplitude parameters, which indicates that it is slightly suboptimal.
Different choices of amplitude-parameter priors have been discussed in~\cite{prix09:_bstat,prix11:_transient},
and a more physical reparametrisation was introduced in~\cite{whelan2014:_ampparams}. 

However, the detector data also contains non-Gaussian disturbances and \emph{artefacts} of instrumental
and environmental origin.
CW searches are mainly affected by so-called 'lines':
narrow-band disturbances that are present for a sizeable fraction of the observation
time~\cite{christensen2010:_S6detchar,coughlin2010:_lineidentification,Aasi2012:_virgochar,
Accadia2012:_noemi,aasi13:_eathS5,aasi2014:_S6detchar}.

Line artefacts are problematic because they can be 'signal-like' in the sense of being more similar to a
CW signal than to Gaussian noise.
Hence, they can cause significant outliers in a CW analysis that is based on the comparison of the signal
model to Gaussian noise only leading to false alarms and therefore to decreased chances of detecting an
actual signal.

Many ad-hoc approaches to mitigate the problem of lines have been developed in the past (see, e.g.,
\cite{christensen2010:_S6detchar,coughlin2010:_lineidentification} and the references in section~II.B of
\cite{keitel2014:_linerobust}).
A Bayesian model-selection approach to line mitigation was first developed in \cite{keitel2014:_linerobust}.
The idea is to use a simple 'signal-like' model for line disturbances in a single detector.
This does not require additional information about the characteristics of GW detectors, but only uses the main
data stream already in use by the standard search methods, specifically the single- and multi-detector
$\F$-statistics.
Thus, it can be viewed as a generalisation of the \emph{$\F$-statistic consistency veto} discussed and used in
\cite{aasi13:_eathS5,aasi2013:_gc-search,behnke2014:_gcmethods}.
This way, \emph{line-robust detection statistics} were derived, referred to in the following as $\OSL$ and
$\OSN$ in the notation of~\cite{keitel2014:_linerobust}.
$\OSL$ is specialised to the test of signals against lines, while $\OSN$ tests signals against a combined
Gaussian noise + lines hypothesis, with a variable transition scale given by a parameter $\Ftho$.
This can be tuned empirically so that $\OSN$ reproduces the performance of the standard multi-detector
$\F$-statistic in Gaussian data, but gives significant improvements in detection probability in line-affected
frequency bands.
We briefly summarise the definition of these statistics in \sref{sec:recap} of this paper.

In this paper, we investigate the behaviour of the line-robust statistics under a set of conditions which
have not been tested in \cite{keitel2014:_linerobust}.
The idea of using a comparison between multi- and single-detector statistics, $\F$ and $\FX$, to distinguish
CW signals from (non-coincident) lines implicitly relies on all detectors having similar sensitivities.
Indeed, for the synthetic tests in \cite{keitel2014:_linerobust} equal noise power spectral densities (PSDs)
$\SnX$ were explicitly assumed, and in the tests on LIGO S5 data the largest deviation between the two
detectors H1 and L1 was \mbox{$\sqrtSnLHO/\sqrtSnLLO \approx 1.7$} (coherent example (\bcoh) in Table~1 of
\cite{keitel2014:_linerobust}).
In addition, all tests so far have been for all-sky searches, averaging out the different antenna patterns of
the individual detectors.

However, very different sensitivities could make signals and lines more difficult to distinguish by this
method, because a signal may yield a significant outlier in only one detector and therefore appear
'line-like' for $\OSL$ and $\OSN$.
This would lead to decreased detection power of the line-robust statistics both in the presence of lines and
in pure Gaussian noise.

In \sref{sec:safety} we will investigate this concern about their \emph{safety} under
these generalised conditions, in the sense that they should never have worse detection probabilities than the
standard $\F$-statistic.
With numerical tests based on \emph{synthetic draws} of $\F$-statistic values, it turns out that this issue
only really affects $\OSL$ and $\OSN$ with a transition-scale parameter $\Ftho$ that is too low.
An optimally-tuned $\OSN$ (in the sense of section~VI.B of \cite{keitel2014:_linerobust}) is found to be safe
under most circumstances of practical relevance, though it can no longer provide improvements over $\F$ in
more extreme cases.

In \sref{sec:sens-weight}, we discuss an attempt to improve upon these statistics.
Using a different amplitude-prior distribution than in \cite{keitel2014:_linerobust}, we obtain new
per-detector sensitivity-weighting factors in the detection statistics, which depend on the noise PSDs, the
amount of data and the sky-location-dependent detector responses.
Synthetic tests show that this extra weighting does recover some of the losses of $\OSL$ and of $\OSN$ tuned
at relatively low $\Ftho$, but that it brings no further improvement compared to an optimally-tuned $\OSN$.

All numerical results in this paper are produced with the same synthesis approach as described in
\cite{prix09:_bstat,prix11:_transient,keitel2014:_linerobust}.
The discussion in this paper is limited to coherent statistics.

\section{Summary of detection statistics}
\label{sec:recap}

Here we give a short introduction to the various detection statistics considered here, which were discussed in
more detail in \cite{keitel2014:_linerobust}.
These are based on comparing three different hypotheses about the observed data: a Gaussian noise hypothesis
$\HypG$, a CW signal hypothesis $\HypS$ and a simple non-coincident 'line' hypothesis $\HypL$.

This section also serves as an introduction to the notation used in this paper.
By $x^X(t)$ we denote a time series of GW strain measured in a detector $X$.
Following the multi-detector notation of~\cite{cutler05:_gen_fstat,prix06:_searc}, boldface indicates a
multi-detector vector, i.e., we write $\dVx(t)$ for the multi-detector data vector with components $x^X(t)$.

\subsection{Posterior probabilities for the different hypotheses}
\label{sec:recap-hyp-prob}

Skipping the derivations given in \cite{keitel2014:_linerobust}, we start from the posterior probabilities.
For the Gaussian-noise hypothesis, \mbox{$\HypG: \dVx(t) = \dVn(t)$}, we have
\begin{equation}
  \label{eq:pHG}
  \prob{\HypG}{\dVx} = \frac{\probI{\HypG}}{\probI{\dVx}}\,\kappa\,\eto{-\frac{1}{2}\scalar{\dVx}{\dVx}} \,,
\end{equation}
with a normalisation constant $\kappa$ and a scalar product between time series defined as
\begin{equation}
  \label{eq:scalarproduct}
  \scalar{\dVx}{\dVy} \equiv \sum_X \frac{1}{\SnX}\int_0^T x^X(t)\,y^X(t)\,dt\,,
\end{equation}
with the single-sided power-spectral densities $\SnX$ assumed as uncorrelated between different detectors $X$
and constant over the (narrow) frequency band of interest.

For the signal hypothesis $\HypS$, a CW waveform $\dVh(t;\Amp,\Dop)$ is determined by four basis functions
$\dVh_\mu(t;\Dop)$, the \emph{amplitude parameters} $\Amp$ and the \emph{phase-evolution parameters} $\Dop$:
\begin{equation}
  \label{eq:Amuhmu}
  \HypS: \dVx(t) = \dVn(t) + \dVh(t;\Amp,\Dop) = \dVn(t) + \Amp^\mu\,\dVh_\mu(t;\Dop)\,,
\end{equation}
which is often referred to as the \emph{JKS factorisation}, first introduced in \cite{jks98:_data}.

For fixed $\Dop$ and an amplitude-parameter prior distribution (as discussed in
\cite{prix09:_bstat,prix11:_transient,keitel2014:_linerobust}, and which we will revisit in
\sref{sec:sens-weight-priors}) of
\begin{equation}
  \label{eq:priorA-rhomax}
  \prob{\{\Amp^\mu\}}{\HypS} = \left\{\begin{array}{ll}
      C & \mathrm{for} \quad h_0^4(\Amp) < \frac{\rhomax^4}{\sqrt{\detM}}\,,\\
      0 & \mathrm{otherwise}\,,
    \end{array}\right.
\end{equation}
with a free cut-off parameter $\rhomax \in (0,\infty)$, the posterior probability is
\begin{equation}
  \label{eq:pHS}
  \prob{\HypS}{\dVx} = \oSG\,\frac{70}{\rhomax^4}\,\prob{\HypG}{\dVx}\,\eto{\F(\dVx)}\,,
\end{equation}
with prior odds $\oSG\equiv \probI{\HypS}/\probI{\HypG}$, and the
well-known multi-detector $\F$-statistic \cite{jks98:_data,cutler05:_gen_fstat} given by 
\begin{equation}
  \label{eq:Fstat}
  2\F(\dVx) \equiv x_\mu\,\M^{\mu\nu}\,x_\nu\,,
\end{equation}
with implicit summation over repeated amplitude indices $\mu$, $\nu$ and the shorthand notations
\begin{equation}
  \label{eq:xmuMmunu}
  x_\mu \equiv \scalar{\dVx}{\dVh_\mu} \quad \mathrm{and} \quad
  \M_{\mu\nu} \equiv \scalar{\dVh_\mu}{\dVh_\nu}\,.
\end{equation}

For the line hypothesis $\HypL$, corresponding to
\mbox{$\HypL^{X} :  x^{X}(t) = n^{X}(t) + h^{X}(t;\Amp^{X})$} in an arbitrary single detector $X$,
a similar derivation leads to
\begin{equation}
  \label{eq:pHL}
  \prob{\HypL}{\dVx} = \frac{70}{\rhomax^4} \, \prob{\HypG}{\dVx} \, \oLG\, \avgX{\rX\,\eto{\FX(x^X)}}\,,
\end{equation}
where we define an average over detectors as
\begin{equation}
  \label{eq:avgX}
  \avgX{Q^X} \equiv \frac{1}{\Ndet} \sum_X Q^X\,,
\end{equation}
and where again the prior-normalisation parameter $\rhomax$ appears.
Through \mbox{$\oLGX \equiv {\probI{\HypL^X}}/{\probI{\HypG^X}}$}
we have \mbox{$\oLG \equiv \sum_X \oLGX$} and, for $\Ndet$ detectors:
\begin{equation}
  \label{eq:rX}
  \rX \equiv \frac{\oLGX}{\oLG / \Ndet}\,,\quad\textrm{such that}\quad \sum_X \rX = \Ndet\,.
\end{equation}

Furthermore, we can combine the (mutually exclusive) hypotheses $\HypG$ and $\HypL$ into an extended noise
hypothesis $\HypN \equiv ( \HypG \OR \HypL )$, with posterior probability
\begin{eqnarray}
 \label{eq:pHN}
  \prob{\HypN}{\dVx} &= \prob{\HypG}{\dVx} + \prob{\HypL}{\dVx} \nonumber \\
                     &= \prob{\HypG}{\dVx} \left( 1 + \frac{70}{\rhomax^4} \,\oLG \,\avgX{\rX \eto{\FX(x^X)}} \right) \,.
\end{eqnarray}

\subsection{Odds ratios}
\label{sec:recap-odds}

These posterior probabilities can be used to compute odds ratios between the different hypotheses.
First, we see from \eref{eq:pHG} and \eref{eq:pHS} that
\begin{equation}
 \label{eq:OSG}
 \OSG(\dVx) \equiv \frac{\prob{\HypS}{\dVx}}{\prob{\HypG}{\dVx}} \propto \eto{\F(\dVx)}\,,
\end{equation}
i.e., this Bayesian approach reproduces the $\F$-statistic as the optimal detection statistic for CW
signals in pure Gaussian noise and under the prior \eref{eq:priorA-rhomax}.

Alternatively, using the posterior probabilities given by \eref{eq:pHS} and \eref{eq:pHL}, we obtain
the posterior signal-versus-line odds as
\clearpage
\begin{equation}
  \label{eq:OSL}
  \OSL(\dVx) \equiv \frac{\prob{\HypS}{\dVx}}{\prob{\HypL}{\dVx}} =
  \oSL \; \frac{ \eto{\F(\dVx)} }{\avgX{\rX\,\eto{\FX(x^X)}}}\,,
\end{equation}
with the prior odds $\oSL \equiv \probI{\HypS}/\probI{\HypL} = \oSG/\oLG$.
Note that the amplitude-prior cut-off $\rhomax$ has disappeared, as we have used the same amplitude
prior on lines and signals.

Finally, using \eref{eq:pHS} and \eref{eq:pHN}, we obtain generalised signal-versus-noise odds
\begin{equation}
  \label{eq:OSN}
  \OSN(\dVx) = \oSN \, \frac{\eto{\F(\dVx)}}
  {(1-\lineprob)\,\eto{\Ftho} + \lineprob\, \avgX{\rX \eto{\FX(x^X)}}}\,,
\end{equation}
where we have rewritten some of the prior parameters in terms of the line probability
\begin{equation}
  \label{eq:lineprob}
  \lineprob \equiv \prob{\HypL}{\HypN} = \frac{\oLG}{1 + \oLG} \in [0,1]
\end{equation}
and a transition-scale parameter
\begin{equation}
  \label{eq:Fth0}
  \Ftho \equiv \ln \frac{\rhomax^4}{70}\,.
\end{equation}
Note that, contrary to the previous odds $\OSG$ and $\OSL$, now the choice of amplitude-prior cut-off
parameter $\rhomax$ will affect the properties of the resulting statistic.
Methods for tuning these free parameters have been discussed in section~VI.B of
\cite{keitel2014:_linerobust}.

\section{Safety of the line-robust statistics for unequal sensitivities}
\label{sec:safety}

Consider a network of two detectors $\detA$ and $\detB$, with $\detA$ being much more sensitive than $\detB$.
There may be CW signals that are strong enough to cause a significant outlier in the single-detector
$\F^\detA$-statistic, but fail to do so in $\F^\detB$, because the signal is still buried in the higher noise
level of detector $\detB$.
For such a network, the multi-detector $\F$-statistic is dominated by the contribution from the more sensitive
detector $\detA$, so that either for a signal or for a strong line in $\detA$, $\F \approx \F^\detA$
holds.
Hence, in this case both an actual astrophysical CW signal and an instrumental line can have very similar
signatures in terms of the set of values \mbox{$\{\F,\FX\}=\{\F,\F^\detA,\F^\detB\}$}.

The line-veto statistic $\OSL$ and line-robust statistic $\OSN$, as given in \eref{eq:OSL} and \eref{eq:OSN},
are intended to suppress 'signal-like' lines and thus they always include a test  of tentative signals
against the line hypothesis.
If, however, weaker signals appear as 'line-like' in terms of $\{\F,\FX\}$, those also receive low odds.
Therefore, it can be expected that these statistics will have problems distinguishing lines from signals in such
unequal-sensitivity cases, losing detection power due to increased false dismissals.

In order to quantify under which conditions a problem may occur, recall the definition of the multi-detector
$\F$-statistic from \eref{eq:Fstat}.
The sensitivity of a detector network is encoded in the antenna-pattern matrix $\M_{\mu\nu}$, which for
high-frequency GWs is given approximately by~\cite{prix:_cfsv2}
\begin{equation}
 \label{eq:Mmunu}
 \M_{\mu\nu} \approx \SinvT
 \left(\matrix{
  A & C & 0 & 0 \cr
  C & B & 0 & 0 \cr
  0 & 0 & A & C \cr
  0 & 0 & C & B
 }\right) \,,
\end{equation}
\clearpage
\noindent and specifically in its determinant
\begin{equation}
 \label{eq:detM}
 \detM \equiv \SinvTf D^2 \,.
\end{equation}
Here, $\Sntot$ is the multi-detector noise PSD,
$\Tdata$ is the effective amount of data
and \mbox{$D \equiv AB-C^2$} quantifies the antenna-pattern-based sensitivity to a particular sky location.
See \ref{sec:antpat-appendix} for full expressions of the \emph{antenna-pattern matrix elements} $A$, $B$,
$C$, $D$.

The corresponding single-detector quantity is
\begin{equation}
 \label{eq:detMX}
 \detMX = \SinvTf \left(D^X\right)^2 \,,
\end{equation}
where $D^X$, through the noise-weighted average from \eref{eq:SFTavg_noiseweighted_X}, depends
quadratically on $(\SnX)^{-1}$ and $\TdataX$.
Example plots for $D^X(\alpha,\delta)$ are also given in \ref{sec:antpat-appendix}.

Thus, for two given detectors, their relative sensitivities are given by their noise PSDs $\SnX$, their
amounts of data $\TdataX$ and the relative sky-position sensitivities $D^X$.
In the following, we will only consider the first and third contribution, since $\TdataX$ enters to the same
power as $\SinvX$ and is therefore equivalent to a corresponding change in that quantity.
We first consider the case of two colocated detectors, for example LIGO H1 and H2 at Hanford (Washington
state), for different $\SnX$ and various noise distributions, in
sections~\ref{sec:safety-gauss}--\ref{sec:safety-H1lines}, and then the case of
non-colocated detectors (LIGO H1 and L1 at Livingston, Louisiana) with different $D^X$ in
\sref{sec:safety-H1L1}.

Here and in \sref{sec:sens-weight-difftests}, $\OSL$ and $\OSN$ use (truncated) 'perfect-knowledge' line
priors:
\mbox{$\oLGX = \max\{ 0.001, \linefrac^X/(1-\linefrac^X) \}$} for a \emph{line-contamination fraction}
$\linefrac^X$, i.e. when a percentage $\linefrac^X$ of noise candidates in detector $X$ comes from lines and
the complement comes from pure Gaussian noise.
All receiver-operating characteristic (ROC) curves -- comparisons of detection probabilities $\pDet$ as a
function of false-alarm probability $\pFA$ -- are based on $\Ndraws=10^7$ each for noise and signal
populations, while the two-dimensional parameter-space exploration plots have $\Ndraws=10^5$ per parameter
combination.

\begin{figure}[t!]
 \begin{minipage}[b]{0.49\textwidth}
  \raggedright (a)\\\vspace*{-0.5cm}
  \includegraphics[width=\textwidth]{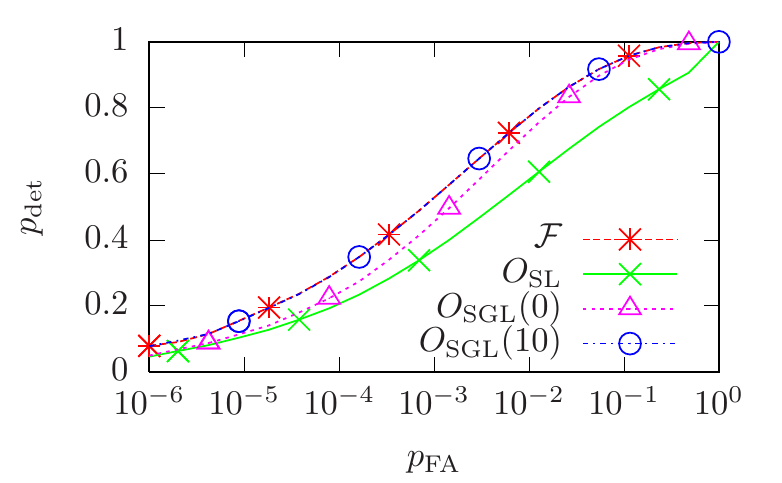}
 \end{minipage}
 \begin{minipage}[b]{0.49\textwidth}
  \raggedright (b)\\\vspace*{-0.5cm}
  \includegraphics[width=\textwidth]{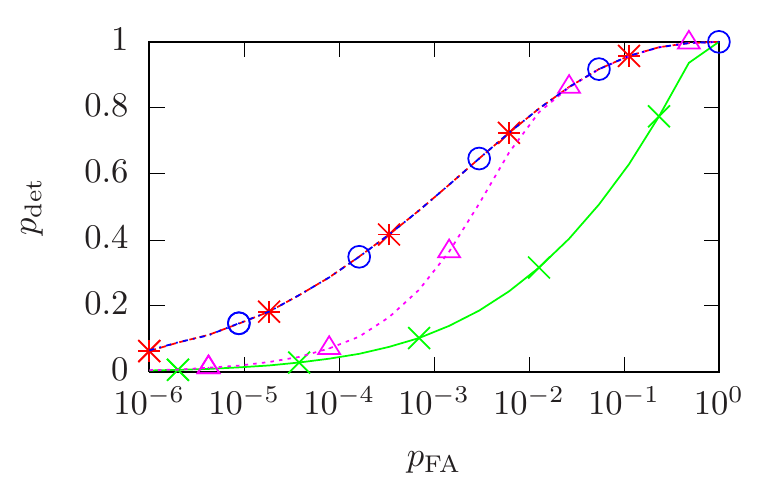}
 \end{minipage}
 \caption{
  \label{fig:safety-rocs-H1H2-gauss}
  Receiver-operating characteristics (ROCs) for signals in pure Gaussian noise:
  Detection probability $\pDet$ as a function of false-alarm probability $\pFA$ for different synthetic
  statistics, for detectors H1 and H2
  and signals with $\snrS=4$ in Gaussian noise without line contamination.
  The panels show relative detector sensitivities of
  \mbox{(a) $\sqrtSnLHOtwo=\sqrtSnLHO$},
  \mbox{(b) $\sqrtSnLHOtwo=10\sqrtSnLHO$}. 
  For $\OSN$, different values of $\Ftho$ are given in brackets.
 }
\end{figure}

\clearpage

\begin{figure}[h!]
 \begin{minipage}[b]{0.49\textwidth}
  \raggedright (a)\\\vspace*{-0.5cm}
  \includegraphics[width=\textwidth]{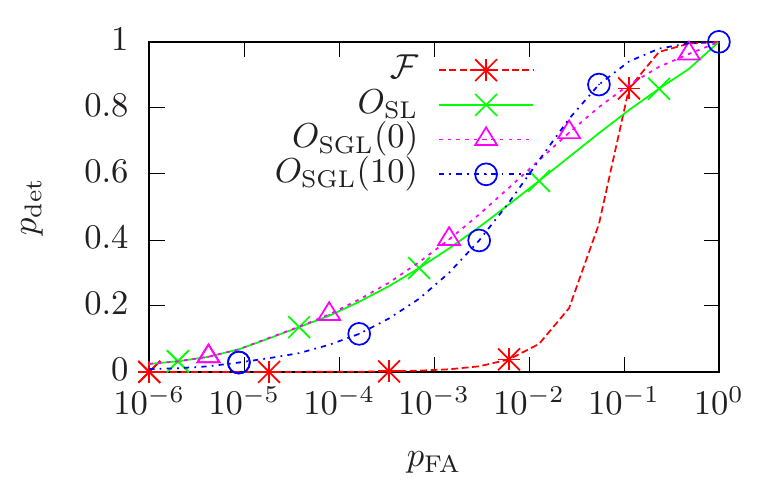}
 \end{minipage}
 \begin{minipage}[b]{0.49\textwidth}
  \raggedright (b)\\\vspace*{-0.5cm}
  \includegraphics[width=\textwidth]{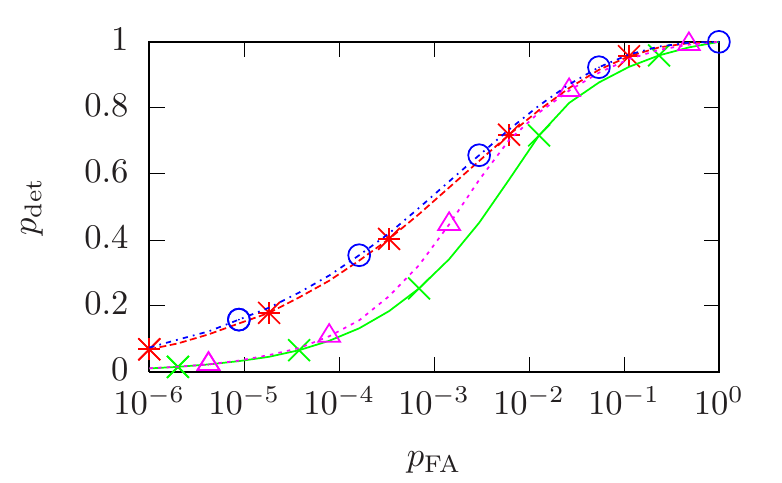}
 \end{minipage}
 \vspace{-0.25cm}
 \caption{
  \label{fig:safety-rocs-H1H2-linesH2}
  Lines in the less sensitive detector:
  ROCs for an H1-H2 network, signals with $\snrS=4$ and
  line-contamination fraction \mbox{$\linefrac^{\LHOtwo}=0.1$}, with $\snrL=6$.
  The panels show relative detector sensitivities of 
  \mbox{(a) $\sqrtSnLHOtwo=\sqrtSnLHO$},
  \mbox{(b) $\sqrtSnLHOtwo=10\sqrtSnLHO$}.
 }
\end{figure}

\begin{figure}[b!]
 \vspace{-0.5\baselineskip}
 \newsavebox{\FigBoxDetProbsHHVarLineSnrWeakLines}
 \sbox{\FigBoxDetProbsHHVarLineSnrWeakLines}{\includegraphics[width=0.5\textwidth]{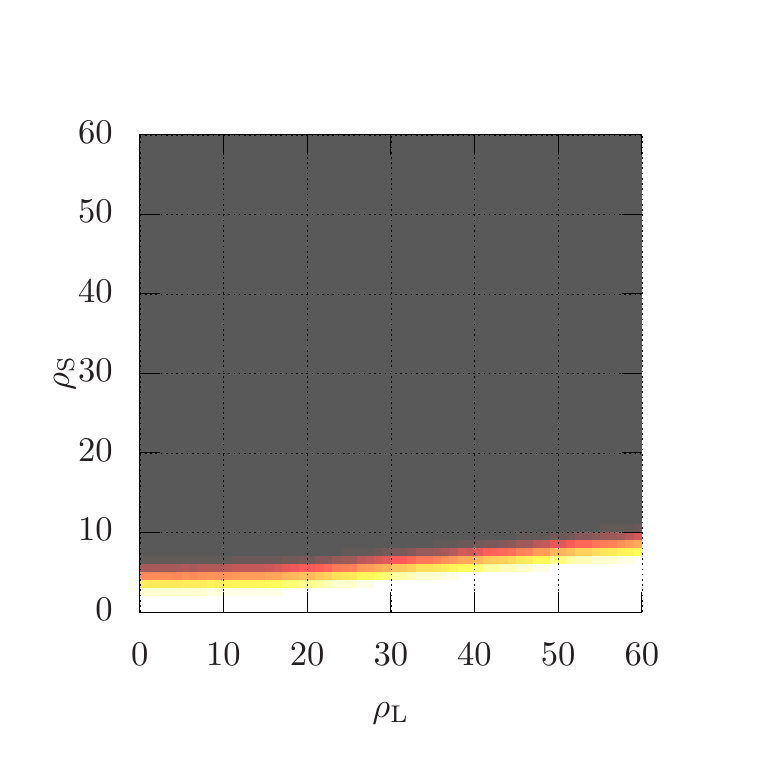}}
 \begin{minipage}[b]{0.5\textwidth}
  \centering $2\F$\\\vspace*{-1.0cm}
  \usebox{\FigBoxDetProbsHHVarLineSnrWeakLines}
 \end{minipage}
 \hspace{-0.5cm}
 \begin{minipage}[b]{0.05\textwidth}
  \includegraphics[height=\ht\FigBoxDetProbsHHVarLineSnrWeakLines]{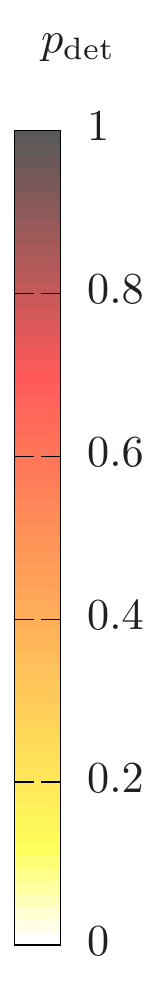} 
  \vspace{-3.4cm} 
 \end{minipage}
 \hspace{0.5cm}
 \begin{minipage}[b]{0.5\textwidth}
  \centering $\OSL$\\\vspace*{-1.0cm} 
  \includegraphics[width=\textwidth]{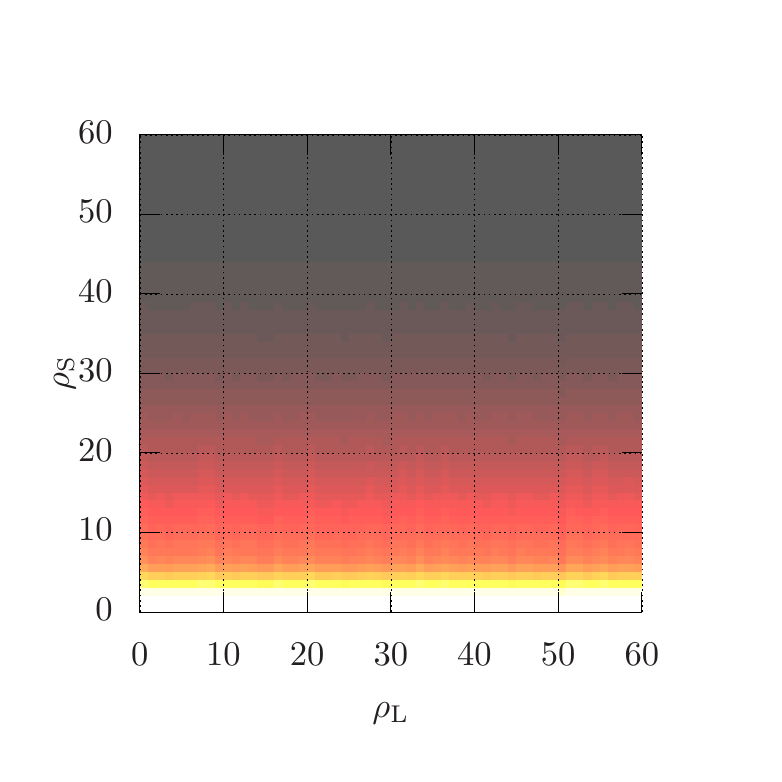}
 \end{minipage}\\[-0.0cm]
 \begin{minipage}[b]{0.5\textwidth}
  \centering \mbox{$\OSNnull$}\\\vspace*{-1.0cm}
  \includegraphics[width=\textwidth]{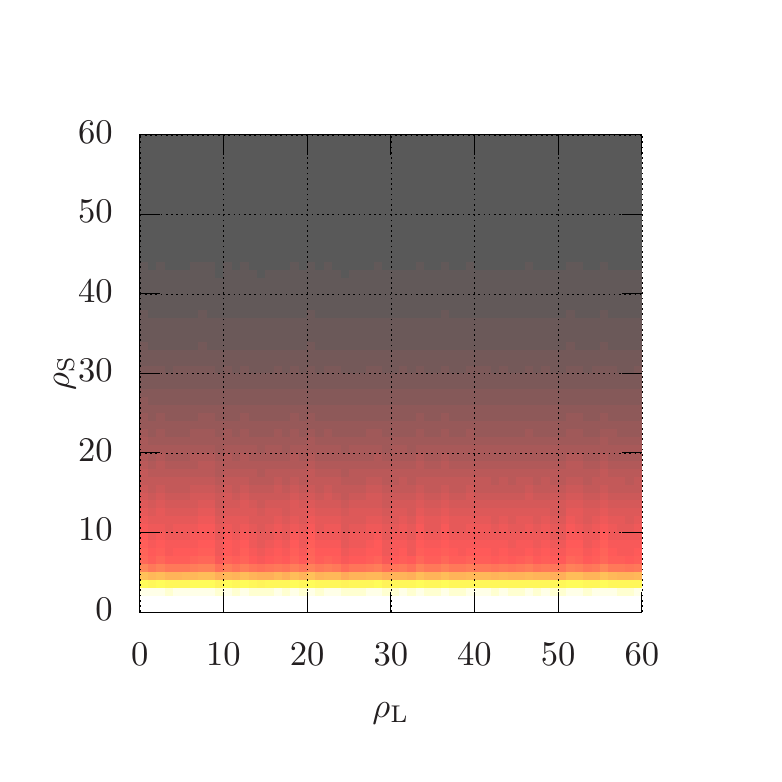}
 \end{minipage}
 \hspace{-0.5cm}
 \begin{minipage}[b]{0.05\textwidth}
  \mbox{} 
 \end{minipage}
 \hspace{0.5cm}
 \begin{minipage}[b]{0.5\textwidth}
  \centering \mbox{$\OSNten$}\\\vspace*{-1.0cm}
  \includegraphics[width=\textwidth]{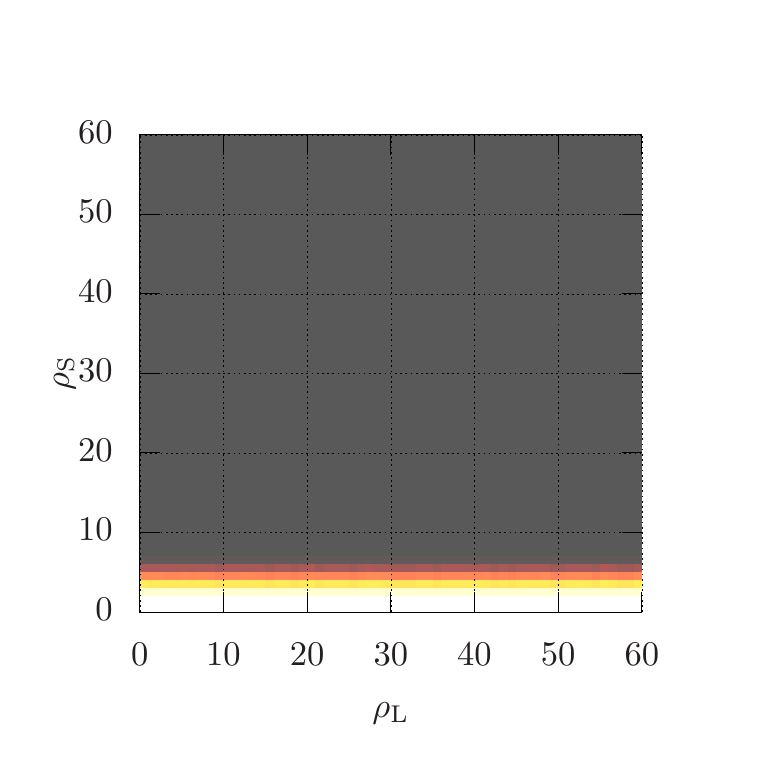}
 \end{minipage}
 \vspace{-1cm}
 \caption{
  \label{fig:safety-H1H2-linesH2-paramspace-varlinesnr-pFA0.001-detprobs}
  Lines in the less sensitive detector:
  detection probabilities $\pDet$ of different statistics,
  for an H1-H2 network with \mbox{$\sqrtSnLHOtwo=10\sqrtSnLHO$},
  as a function of line SNR $\snrL$ and signal SNR $\snrS$
  at fixed line contamination \mbox{$\linefrac^{\LHOtwo}=0.1$} and false-alarm probability \mbox{$\pFA=0.001$}.
 }
\end{figure}

\clearpage

\subsection{Gaussian noise}
\label{sec:safety-gauss}

ROC curves for synthetic draws from a signal population with fixed signal-to-noise ratio (SNR) of $\snrS=4$
and pure Gaussian noise without line contamination are shown in \fref{fig:safety-rocs-H1H2-gauss}.
In panel (a), both detectors have the same PSD, \mbox{$\sqrtSnLHOtwo=\sqrtSnLHO$}.
The results are very similar to those from \cite{keitel2014:_linerobust}, where an H1-L1 network was used with
identical parameters otherwise.
$\OSN$ with an optimal tuning of $\Ftho=10$ reproduces the detection probabilities of the $\F$-statistic,
while $\OSL$ and \mbox{$\OSNnull$} have up to 20\% lower $\pDet$.

For extremely unequal sensitivities, \mbox{$\sqrtSnLHOtwo=10\sqrtSnLHO$}, as shown in panel (b), the losses of
$\OSL$ and \mbox{$\OSNnull$} become more pronounced, due to increased false dismissals of 'line-like'
signals lowering $\pDet$ even in the absence of lines.
However, \mbox{$\OSNten$} is \emph{not} affected, because it still lends more weight to the Gaussian-noise
hypothesis over the line hypothesis, thus being less likely to confuse signals with lines.
This shows that the 'safety'-tuning approach of \cite{keitel2014:_linerobust} still works even in this
extreme example.

\subsection{Lines in the less sensitive detector}
\label{sec:safety-H2lines}

For the extreme case of \mbox{$\sqrtSnLHOtwo=10\sqrtSnLHO$}, lines in the less sensitive detector H2 are
already strongly suppressed in the multi-detector $\F$-statistic, resulting in a population of $\F$-statistic
noise candidates that is quite similar to the case of pure Gaussian noise up to high line SNRs $\snrL$.

ROCs for this case, with a line-contamination fraction \mbox{$\linefrac^{\LHOtwo}=0.1$} and line strength
$\snrL=6$, are shown in panel (b) of \fref{fig:safety-rocs-H1H2-linesH2}, compared to the equal-sensitivity
case and identical line parameters in panel (a).
Whereas the lines in (a) have a strong effect on $\pDet$ for the various statistics (again similar to the
results in \cite{keitel2014:_linerobust}), (b) is closer to the Gaussian case (panel (b) of
\fref{fig:safety-rocs-H1H2-gauss}), with losses in sensitivity for $\OSL$ and $\OSNnull$, while
\mbox{$\OSNten$} marginally outperforms the $\F$-statistic.

In \fref{fig:safety-H1H2-linesH2-paramspace-varlinesnr-pFA0.001-detprobs}, detection probabilities
at fixed \mbox{$\pFA=0.001$} are shown for the statistics $2\F$, $\OSL$, \mbox{$\OSNnull$} and
\mbox{$\OSNten$} over a wide range in $\snrS$ and $\snrL$ and with a line contamination of
\mbox{$\linefrac^{\LHOtwo}=0.1$}.

The results show a very weak dependence on $\snrL$.
Both the $\F$-statistic and the line-robust \mbox{$\OSNten$} perform, over most of the range, as well as in
Gaussian noise and for equal sensitivities.
The tuned line-robust statistic \mbox{$\OSNten$} outperforms the $\F$-statistic only at very high $\snrL$.

Meanwhile, $\OSL$ and \mbox{$\OSNnull$} show a mostly $\snrL$-independent deficiency in detection power, only
approaching \mbox{$\pDet=1$} for extremely high $\snrS$ of 40 or higher.

\subsection{Lines in the more sensitive detector}
\label{sec:safety-H1lines}

\begin{figure}[t!]
 \begin{minipage}[b]{0.49\textwidth}
  \raggedright (a)\\\vspace*{-0.5cm}
  \includegraphics[width=\textwidth]{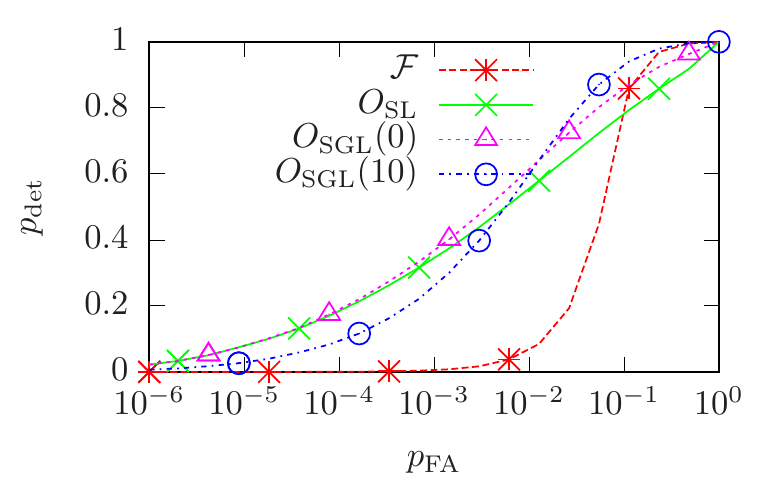}
 \end{minipage}
 \begin{minipage}[b]{0.49\textwidth}
  \raggedright (b)\\\vspace*{-0.5cm}
  \includegraphics[width=\textwidth]{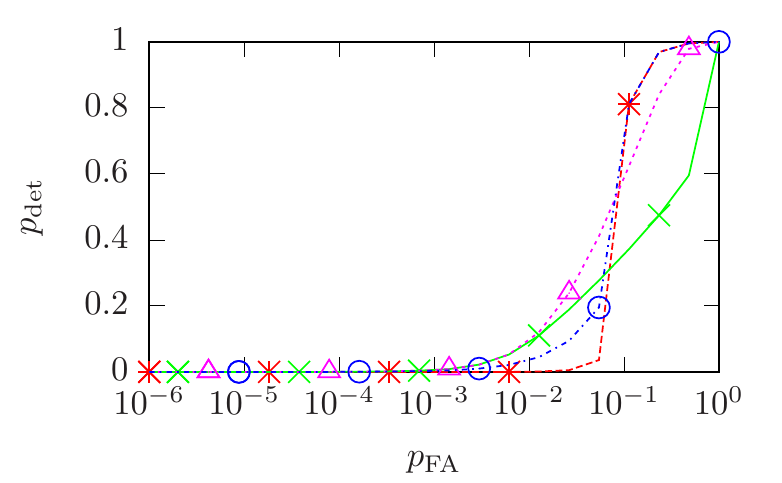}
 \end{minipage}
 \vspace{-0.25cm}
 \caption{
  \label{fig:safety-rocs-H1H2-linesH1}
  Lines in the more sensitive detector: ROCs for an H1-H2 network, signals with $\snrS=4$ and
  line-contamination fraction \mbox{$\linefrac^{\LHO}=0.1$}, with $\snrL=6$.
  The panels show relative detector sensitivities of 
  \mbox{(a) $\sqrtSnLHOtwo=\sqrtSnLHO$},
  \mbox{(b) $\sqrtSnLHOtwo=10\sqrtSnLHO$}.
 }
\end{figure}

Next, we consider the case where again \mbox{$\sqrtSnLHOtwo=10\sqrtSnLHO$}, but now there is a
line-contamination fraction in the more sensitive detector of $\linefrac^{\LHO}=0.1$.
The ROCs in \fref{fig:safety-rocs-H1H2-linesH1} contrast this case (panel b) with the
equal-sensitivity case \mbox{$\sqrtSnLHOtwo=\sqrtSnLHO$} (panel a), both for signals with $\snrS=4$ and lines
with $\snrL=6$.
As found before in \cite{keitel2014:_linerobust}, all variants of the line-robust statistics
give large improvements over the $\F$-statistic in the equal-sensitivity case.
However, in panel (b) most of these improvements disappear, though \mbox{$\OSNten$} is still safe compared to
$2\F$ at all $\pFA$.

\clearpage

\begin{figure}[t!]
 \newsavebox{\FigBoxDetProbsHHVarLineSnrStrongLines}
 \sbox{\FigBoxDetProbsHHVarLineSnrStrongLines}{\includegraphics[width=0.5\textwidth]{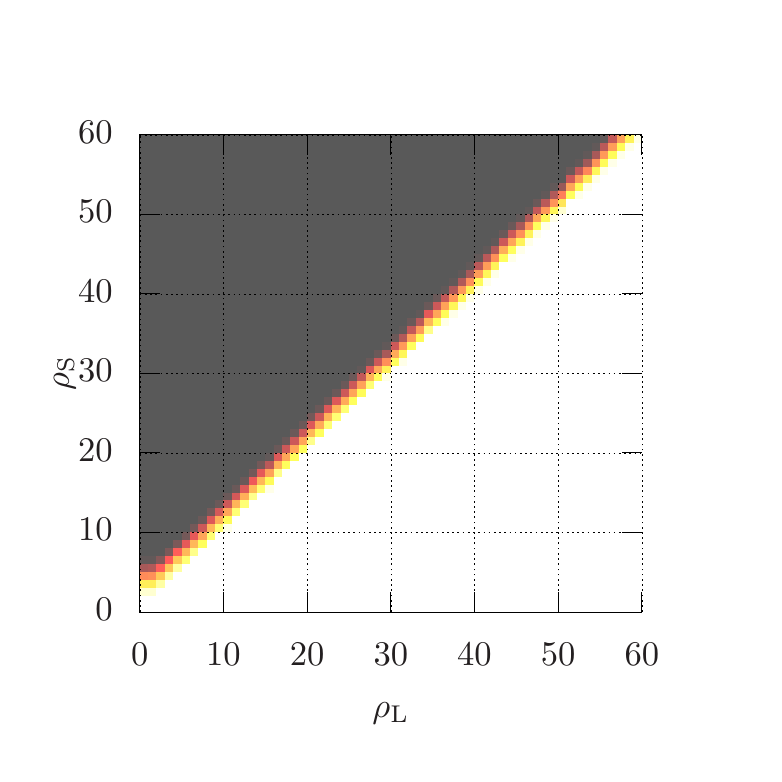}}
 \begin{minipage}[b]{0.5\textwidth}
  \centering $2\F$\\\vspace*{-1.0cm}
  \usebox{\FigBoxDetProbsHHVarLineSnrStrongLines}
 \end{minipage}
 \hspace{-0.5cm}
 \begin{minipage}[b]{0.05\textwidth}
  \includegraphics[height=\ht\FigBoxDetProbsHHVarLineSnrStrongLines]{synth_detprobs_colorbar} 
  \vspace{-3.4cm} 
 \end{minipage}
 \hspace{0.5cm}
 \begin{minipage}[b]{0.5\textwidth}
  \centering $\OSL$\\\vspace*{-1.0cm}
  \includegraphics[width=\textwidth]{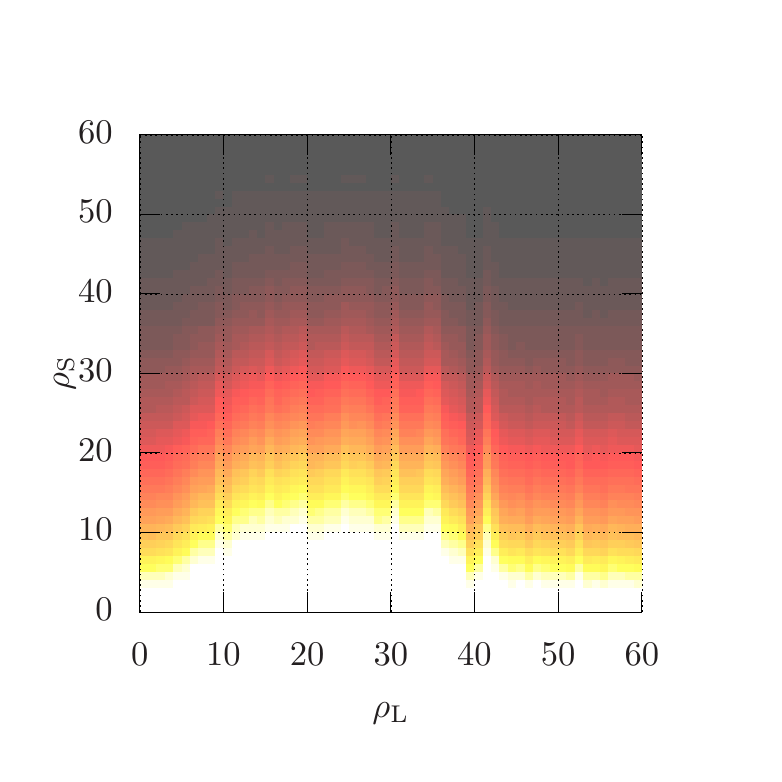}
 \end{minipage}\\[-0.0cm]
 \begin{minipage}[b]{0.5\textwidth}
  \centering \mbox{$\OSNnull$}\\\vspace*{-1.0cm}
  \includegraphics[width=\textwidth]{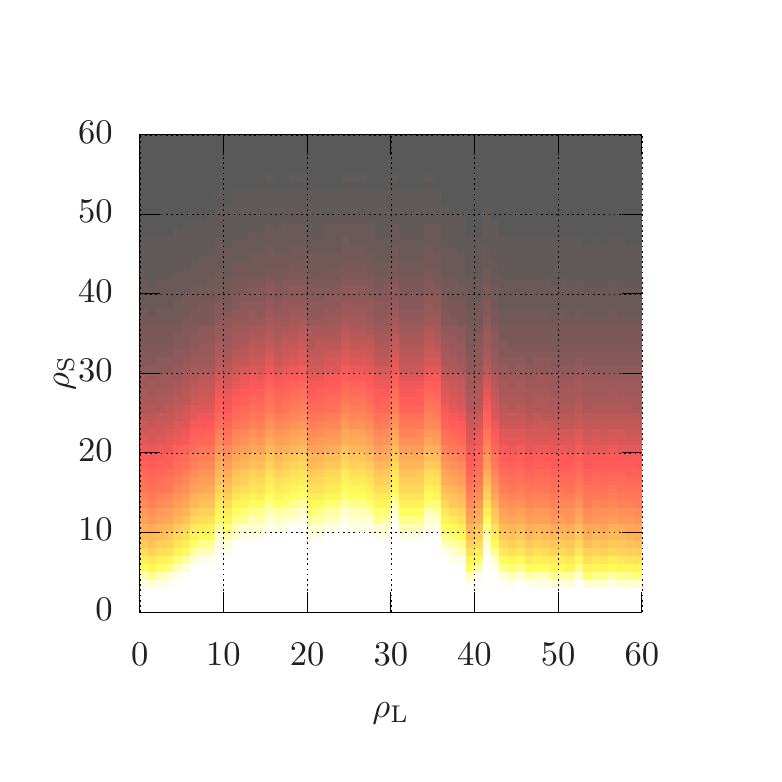}
 \end{minipage}
 \hspace{-0.5cm}
 \begin{minipage}[b]{0.05\textwidth}
  \mbox{} 
 \end{minipage}
 \hspace{0.5cm}
 \begin{minipage}[b]{0.5\textwidth}
  \centering \mbox{$\OSNten$}\\\vspace*{-1.0cm}
  \includegraphics[width=\textwidth]{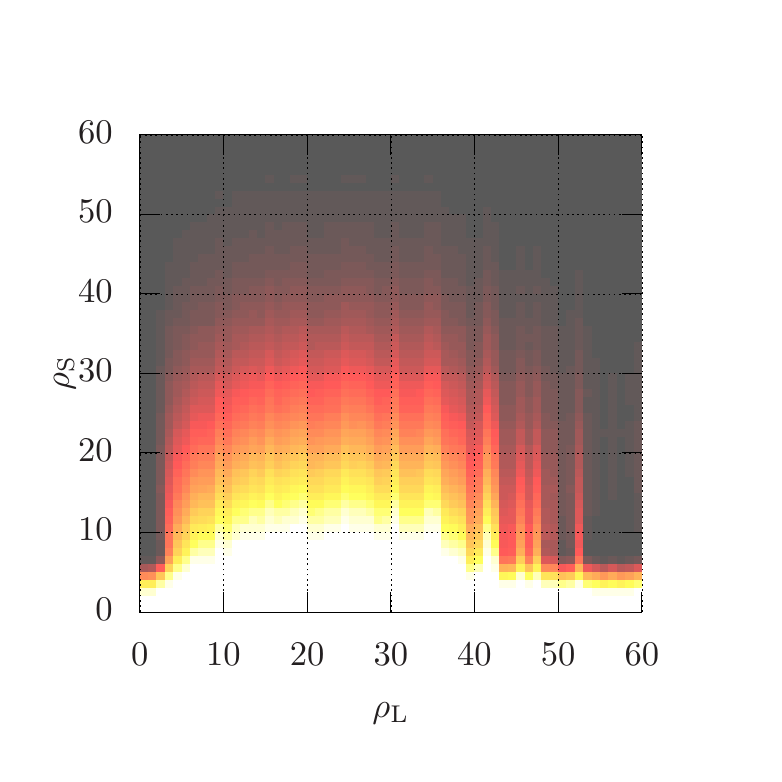}
 \end{minipage}
 \vspace{-1.0cm}
 \caption{
  \label{fig:safety-H1H2-linesH1-paramspace-varlinesnr-pFA0.001-detprobs}
  Lines in the more sensitive detector:
  detection probabilities $\pDet$ of different statistics,
  for an H1-H2 network with \mbox{$\sqrtSnLHOtwo=10\sqrtSnLHO$},
  as a function of line SNR $\snrL$ and signal SNR $\snrS$
  at fixed line contamination \mbox{$\linefrac^{\LHO}=0.1$} and false-alarm probability \mbox{$\pFA=0.001$}.
 }
\end{figure}

\clearpage

\begin{figure}[t!]
  \newsavebox{\FigBoxDetProbDiffsHHVarLineSnrStrongLinesFOSN}
  \sbox{\FigBoxDetProbDiffsHHVarLineSnrStrongLinesFOSN}{\includegraphics[width=0.5\textwidth]{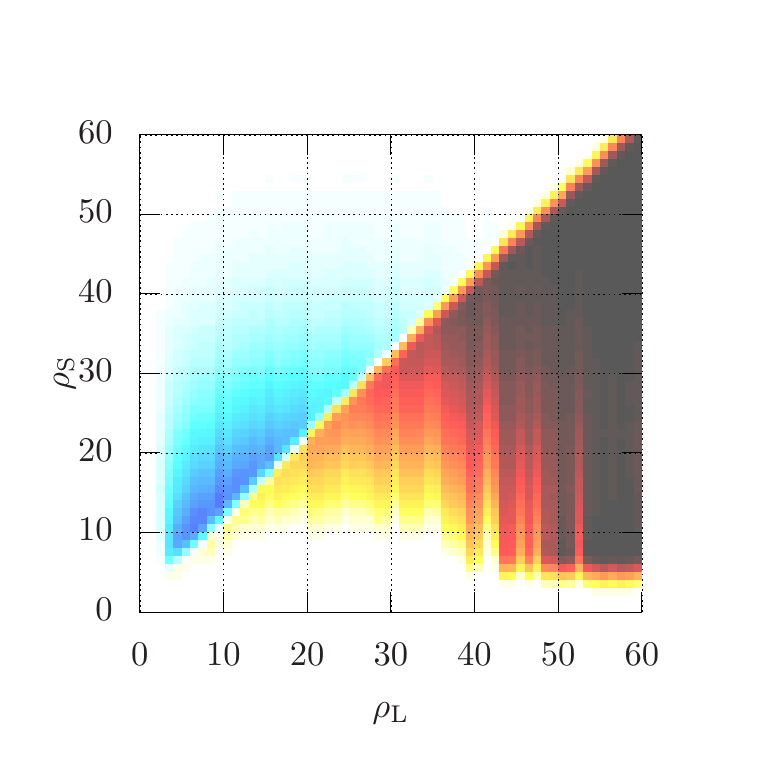}}
   \hspace{-1cm}
  \begin{minipage}[b]{0.5\textwidth}
   \hphantom{xxxxxxxxxxxxxxxx}
  \end{minipage}
  \begin{minipage}[b]{0.45\textwidth}
   \usebox{\FigBoxDetProbDiffsHHVarLineSnrStrongLinesFOSN}
  \end{minipage}
  \begin{minipage}[b]{0.05\textwidth}
   \includegraphics[height=\ht\FigBoxDetProbDiffsHHVarLineSnrStrongLinesFOSN]{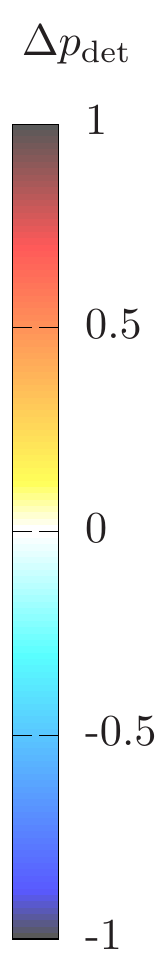} 
  \end{minipage}
  \caption{
   \label{fig:safety-H1H2-linesH1-paramspace-varlinesnr-pFA0.001-detprobdiff-2F-OSN10}
   Lines in the more sensitive detector:
   Detection probability difference \mbox{$\Delta\pDet=\pDet(\OSN)-\pDet(2\F)$} with \mbox{$\Ftho=10$},
   for an H1-H2 network with \mbox{$\sqrtSnLHOtwo=10\sqrtSnLHO$},
   as a function of line SNR $\snrL$ and signal SNR $\snrS$,
   at fixed \mbox{$\linefrac^{\LHO}=0.1$} and \mbox{$\pFA=0.001$}.
  }
\end{figure}

A systematic investigation of detection probabilities over a range of $\snrS$ and $\snrL$ is shown in
\fref{fig:safety-H1H2-linesH1-paramspace-varlinesnr-pFA0.001-detprobs}, at fixed \mbox{$\pFA=0.001$}.
This shows that \mbox{$\OSNten$} performs as well as the $\F$-statistic at very low $\snrL$, where the noise
is still almost Gaussian, for all $\snrS$.
This range also includes the example from \fref{fig:safety-rocs-H1H2-linesH1}.
At very high $\snrL$, \mbox{$\OSNten$} performs much better than $\F$ and almost as well as in the
low-$\snrL$ regime.

In parts of the main region of interest, namely for intermediate to high $\snrL$, \mbox{$\OSNten$} performs
worse than the $\F$-statistic when \mbox{$\snrS>\snrL$}.
However, this is compensated by regions where \mbox{$\OSNten$} performs better than $2\F$, namely for
\mbox{$\snrS<\snrL$}.

A direct comparison of $\pDet$ between \mbox{$\OSNten$} and $2\F$ is shown in
\fref{fig:safety-H1H2-linesH1-paramspace-varlinesnr-pFA0.001-detprobdiff-2F-OSN10}.
Among the results presented in this section, this plot gives the clearest picture of the potential problem
related to the simple line hypothesis introduced in \cite{keitel2014:_linerobust}:
whereas the line-robust statistics improve over the $\F$-statistic by suppressing 'signal-like lines', the
approach can fail when there are 'line-like signals' in the data because of a much less sensitive detector
not picking up the signal.

Typical CW searches with ground-based detectors operate in the regime of low signal SNRs $\snrS$, so that the
unsafe region for \mbox{$\OSNten$} is likely of little practical relevance.
On the other hand, gains from $\OSN$ in this case are limited to the low-$\snrS$ and high-$\snrL$ range, which
is more likely to be relevant in real data.

\clearpage

\subsection{Sky-location dependence}
\label{sec:safety-H1L1}

For differently oriented detectors, their antenna patterns lead to sky-location-dependent sensitivity
differences, through the determinant factors $D^X$ \eref{eq:detMX}.
However, these differences are quite small, and they partially average out over longer observation times --
as visualised in figures~\ref{fig:antpatD_allsky_12h} and \ref{fig:antpatD_allsky_24h}.
For example, the maximum ratio of antenna-pattern determinants between the H1 and L1 detectors for a 12-hour
observation time 
at any sky location is \mbox{$D^{\LLO}/D^{\LHO}\approx5.84$}.
For a 24-hour observation, this decreases to a maximum ratio of \mbox{$D^{\LLO}/D^{\LHO}\approx2.67$}.

In comparing this discrepancy in sensitivities to those considered in the preceding tests, the scale of $D^X$
ratios can be translated to an equivalent scale of square-roots of the noise PSDs.
Through the noise-weights of \eref{eq:noiseweights}, $D^X$ is proportional to $(\SnX)^{-2}$, so that
these maximum ratios would correspond to equivalent square-root-PSD ratios \mbox{$\sqrtSnLLO/\sqrtSnLHO$} of
only about 1.55 and 1.28, respectively.

Hence, we expect only a very small effect on detection probabilities from this most extreme case of different
antenna patterns, and even less for other sky locations or for all-sky searches.
Synthetic-ROC tests at 'worst-case' sky positions have confirmed this, resulting in no significant losses in
detection power and no safety concerns for either $\OSN$ or $\OSL$.

\section{Sensitivity-weighted detection statistics}
\label{sec:sens-weight}

In this section, we describe and test an idea for improving the line-robust statistics in the case of
detectors with different sensitivities.
The idea is to re-weight the contribution of each detector in the denominators of $\OSL$ and $\OSN$ with a
factor corresponding to its respective sensitivity, including its PSD, the amount of data and sky-location
dependence -- as seen in \eref{eq:detMX}.
By \emph{down-weighting} contributions with higher sensitivity, this should intuitively decrease the chance of
considering candidates with unequal $\FX$-statistics as lines, thus decreasing the risk of false dismissals.
A simple approach for including such a sensitivity weighting consists of changing the amplitude-parameter
prior used to derive the likelihood for $\HypL$ \eref{eq:pHL}.

\subsection{More on amplitude priors}
\label{sec:sens-weight-priors}

The derivation of the posterior probability for the signal hypothesis $\HypS$ \eref{eq:pHS} and $\HypL$
\eref{eq:pHL} contains an integral for the marginalisation over amplitude parameters $\Amp$, namely
\begin{equation}
 \prob{\dVx}{\HypS} = \int \prob{\dVx}{\HypS,\Amp}\,\prob{\Amp}{\HypS}\,d\Amp \,. \label{eq:likeli_HSmarg}
\end{equation}
As discussed in \cite{prix09:_bstat}, this integral cannot be solved analytically for general
parametrisations of $\Amp$ and prior distributions $\prob{\Amp}{\HypS}$.
However, as demonstrated in \cite{prix09:_bstat,prix11:_transient}, it becomes a simple Gaussian integral for
a uniform prior in the four amplitude parameters $\Amp^\mu$ in the 'JKS factorisation' \eref{eq:Amuhmu}.

Such a uniform prior would be 'improper' (non-normalisable), unless we introduce a cut-off.
One possibility is the '$\F$-statistic prior' \eref{eq:priorA-rhomax} introduced in \cite{prix11:_transient}
and adapted in \cite{keitel2014:_linerobust}, corresponding to a cut-off in a SNR-like quantity
\mbox{$\widehat{\snr} \equiv h_0 \detM^{1/8}$}.

However, \cite{prix09:_bstat} originally used a different prior distribution, placing a fixed cut-off
\mbox{$\hmax \in (0,\infty)$} on the signal-strength parameter $h_0$:
\clearpage
\begin{equation}
  \label{eq:sens-weight-priorA-hmax}
  \prob{\{\Amp^\mu\}}{\HypSdetM} = \left\{\begin{array}{ll}
      C & \mathrm{for} \quad h_0(\Amp) < \hmax \,,\\
      0 & \mathrm{otherwise}\,.
    \end{array}\right.
\end{equation}
Here, we use $\HypSdetM$ as a shorthand for this signal hypothesis with modified amplitude prior.

This variant was discarded in \cite{prix11:_transient} due to poor performance of the resulting detection
statistic on medium-duration 'transient CW' signals.
However, for standard CW signals, the effect of such a prior has not been explicitly analysed yet, especially
not in the context of comparing several detectors for robustness against line artefacts.

\subsection{Sensitivity weighting for signals in pure Gaussian noise?}
\label{sec:sens-weight-sig-gauss}

Before considering lines, it clarifies matters to first investigate the effect of these prior choices in the
simpler case of CW signals in pure Gaussian noise.
The difference between the two prior choices is that \eref{eq:priorA-rhomax} results in the
signal-hypothesis posterior \eref{eq:pHS}
\begin{equation}
 \prob{\HypS}{\dVx} = \oSG\,\cF^{-1}\,\prob{\HypG}{\dVx}\,\eto{\F(\dVx)}
\end{equation}
and in signal-to-Gaussian-noise odds \eref{eq:OSG} of
$\OSG(\dVx) \propto \eto{\F(\dVx)}$,
while \eref{eq:sens-weight-priorA-hmax} leads to a signal-hypothesis posterior
\begin{equation}
 \prob{\HypSdetM}{\dVx} = \oSdetMG \, \frac{70}{\hmax^4} \, \prob{\HypG}{\dVx} \, \detM^{-1/2} \,
\eto{\F(\dVx)}
\end{equation}
and odds of
\begin{equation}
 \label{eq:OSGdetM_initial}
 \OSdetMG(\dVx) \equiv \frac{\prob{\HypSdetM}{\dVx}}{\prob{\HypG}{\dVx}}
                \propto \detM^{-1/2} \, \eto{\F(\dVx)} \,.
\end{equation}
As discussed before, the antenna-pattern determinant $\detM$ is a measure of the overall sensitivity of a
network of detectors.
We therefore refer, in the following, to any statistic derived from the prior
\eref{eq:sens-weight-priorA-hmax}, so that it has an explicit factor of $\detM$ in the odds, as a
\emph{sensitivity-weighted statistic}.

Inserting the explicit expression from \eref{eq:detM} yields
\begin{equation}
 \OSdetMG(\dVx) \propto \eto{\F(\dVx)} \, \STinvsq D^{-1} \,,
\end{equation}
demonstrating that any candidate coming from a particularly good set of data (low $\Sntot$, large $\Tdata$), or
from a point on the sky where the detector is most sensitive over the observation time (large $D$), is
actually \emph{down-weighted}.
Thus, intuitively this statistic should be worse than the pure $\F$-statistic, and hence we will not use
$\HypSdetM$ instead of $\HypS$.

\subsection{Sensitivity-weighted line-veto statistic}
\label{sec:sens-weight-line-veto}

On the other hand, the effect of down-weighting outliers from more sensitive data could be useful in the case
of line-vetoing.
Consider again two detectors, one much more sensitive than the other (e.g., having lower $\SnX$), and a signal
that is strong enough to produce an elevated $\FX$-statistic in the better detector, but not strong enough to
be seen in the other one.
This signal is likely to trigger the simple line hypothesis $\HypL$ from \cite{keitel2014:_linerobust}, as the
signature is similar to that of a line in one of two equally sensitive detectors.

When we introduce sensitivity weighting in a modified line hypothesis, an outlier in the
more sensitive detector will be considered less likely to come from a line.
Such a reduction of false positives for the line hypothesis should then lead to less false dismissals of
signals by the signal-versus-line odds.

Hence, we define a sensitivity-weighted single-detector line hypothesis $\HypLdetM^X$ that uses the
single-detector version of the prior from \eref{eq:sens-weight-priorA-hmax}:
\begin{equation}
  \label{eq:sens-weight-priorA-hmax-L}
  \prob{\{\Amp^{X\mu}\}}{\HypLdetM^X} = \left\{\begin{array}{ll}
      C & \mathrm{for} \quad h_0(\Amp^X) < \hmax^X \,,\\
      0 & \mathrm{otherwise}\,.
    \end{array}\right.
\end{equation}
The line hypothesis $\HypLdetM$ for a non-coincident line in any detector is constructed as in
equation~(19) of \cite{keitel2014:_linerobust}.
\emph{Sensitivity-weighted line-veto odds} are then given by the unweighted signal hypothesis $\HypS$
and the weighted $\HypLdetM$:
\begin{equation}
 \hspace{-1cm}
 \label{eq:sens-weight-OSLdetM_rhohX}
 \OSLdetM(\dVx) \equiv \frac{\prob{\HypS}{\dVx}}{\prob{\HypLdetM}{\dVx}}
                = \oSLdetM \; \frac{ \rhomax^{\,-4} \eto{\F(\dVx)} }
                  { \avgX{ \left(\hmax^X\right)^{-4} \rX  \detMX^{-1/2} \, \eto{\FX(x^X)} } } \,,
\end{equation}
where $\rX$ are the line-prior weights defined in \eref{eq:rX}.

This expression contains various prior-cut-off parameters: $\rhomax$ from $\HypS$ and a set of $\{\hmax^X\}$
from $\HypLdetM^X$.
In \cite{keitel2014:_linerobust}, \mbox{$\rhomaxS=\rhomaxL=\rhomaxLX$} was assumed for all $X$, so that
any such parameters cancelled out.
Here, we first assume \mbox{$\hmax^X = \hmax$}, again for all $X$, justified by the general absence of
detailed physical knowledge about different line-strength populations in different detectors.
This reduces the effect of the cut-offs to a common pre-factor to the odds:
\begin{equation}
  \label{eq:sens-weight-OSLdetM_rhoh}
  \OSLdetM(\dVx) = \oSLdetM \; \left( \frac{\hmax}{\rhomax} \right)^4
                   \frac{ \eto{\F(\dVx)} } { \avgX{ \rX  \detMX^{-1/2} \, \eto{\FX(x^X)} } } \,.
\end{equation}
As the only requirement for these cut-offs is that they should be large enough for the marginalisation
integral to become Gaussian, we can choose to keep only one of them as a free parameter and to fix the ratio
between them.
For convenience of notation, we define
\begin{equation}
 \label{eq:sens-weight-averagedetM}
 \avD \equiv \left(\frac{\rhomax}{\hmax}\right)^8 = \const.
\end{equation}
Pulling $\avD$ into the denominator and defining a per-detector relative \emph{sensitivity-weighting factor}
\begin{equation}
 \label{eq:sens-weight-weightfactor}
 \sweightX \equiv \sqrt{\frac{\avD}{\detMX}} \,,
\end{equation}
the weighted line-veto statistic \eref{eq:sens-weight-OSLdetM_rhohX} becomes
\begin{equation}
 \hspace{-2cm}
 \label{eq:sens-weight-OSLdetM}
 \OSLdetM(\dVx) = \oSLdetM \; \frac{ \eto{\F(\dVx)} } { \avgX{ \rX \sweightX \eto{\FX(x^X)} } } \,.
\end{equation}

\subsection{Sensitivity-weighted line-robust statistic}
\label{sec:sens-weight-line-robust}

Furthermore, we construct an extended noise hypothesis, in analogy to equation~(32)
of \cite{keitel2014:_linerobust}, but this time using the modified line-amplitude prior from
\eref{eq:sens-weight-priorA-hmax-L}:
\begin{equation}
 \HypNdetM : \left( \HypG \OR \HypLdetM \right) \,.
\end{equation}
\clearpage
\noindent This gives the \emph{sensitivity-weighted line-robust odds} as an analogue to $\OSN$ of
\eref{eq:OSN}:
\begin{equation}
 \hspace{-2cm}
 \label{eq:sens-weight-OSNdetM}
 \OSNdetM(\dVx) \equiv \frac{\prob{\HypS}{\dVx}}{\prob{\HypNdetM}{\dVx}}
                = \oSNdetM \; \frac{ \eto{\F(\dVx)} } { (1-\lineprob) \, \eto{\Ftho} + \lineprob \,
                  \avgX{ \rX \sweightX \eto{\FX(x^X)} } } \,.
\end{equation}
Here, the transition-scale parameter $\Ftho$ is unchanged from its definition in
\eref{eq:Fth0}, as only the per-detector terms are modified.

\begin{figure}[t!]
 \begin{minipage}[b]{0.49\textwidth}
  \raggedright (a)\\\vspace*{-0.5cm}
  \includegraphics[width=\textwidth]{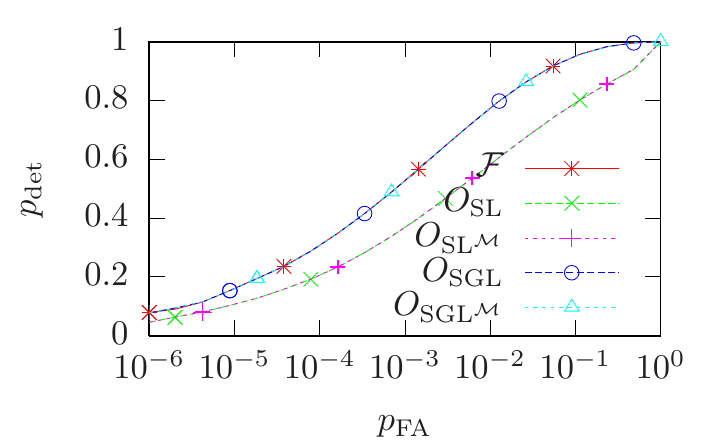} \\
  \raggedright (c)\\\vspace*{-0.5cm}
  \includegraphics[width=\textwidth]{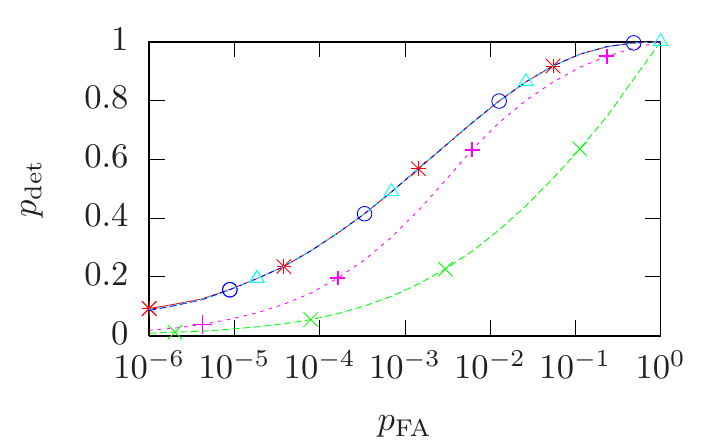} 
 \end{minipage}
 \begin{minipage}[b]{0.49\textwidth}
  \raggedright (b)\\\vspace*{-0.5cm}
  \includegraphics[width=\textwidth]{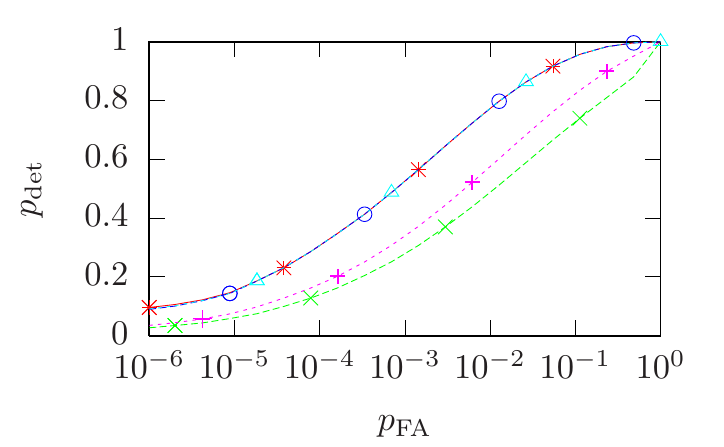} \\
  \raggedright (d)\\\vspace*{-0.5cm}
  \includegraphics[width=\textwidth]{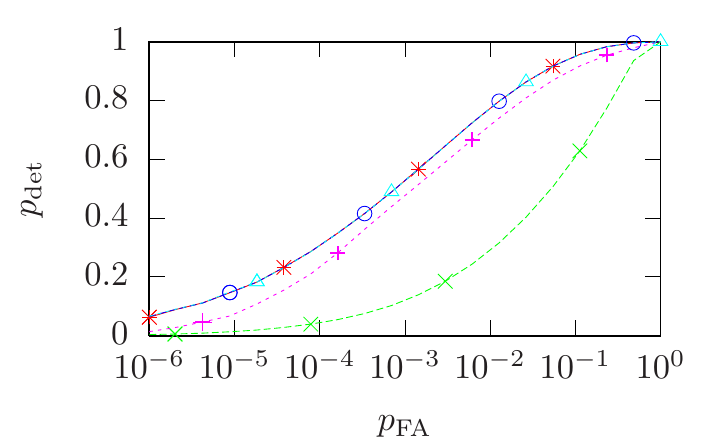} 
 \end{minipage}
 \caption{
  \label{fig:sens-weight-improvement-rocs-H1H2-gauss}
  ROCs for signals with $\snrS=4$ in pure Gaussian noise.
  The panels show relative detector sensitivities of
  \mbox{(a) $\sqrtSnLHOtwo=\sqrtSnLHO$},
  \mbox{(b) $\sqrtSnLHOtwo=2\sqrtSnLHO$},
  \mbox{(c) $\sqrtSnLHOtwo=5\sqrtSnLHO$},
  \mbox{(d) $\sqrtSnLHOtwo=10\sqrtSnLHO$}.
  $\OSN$ and $\OSNdetM$ both use \mbox{$\Ftho=10$}.
 }
\end{figure}

Comparing the relative scale between the constant term and the per-detector terms, a change in
sensitivity-weighting factors $\sweightX$ would only be compensated by a logarithmic change in $\Ftho$.
If we further choose the free parameter $\avD$ as similar to typical $\detMX$ values, all denominator terms in
\eref{eq:sens-weight-OSNdetM} become similar in scale to those in \eref{eq:OSN}.
In principle, $\avD$ should be a constant over the whole parameter space, but as the sky-dependent variations
in $D^X$, and thus $\detMX$, are rather small, we choose a particularly simple prescription:
\begin{equation}
 \avD(\alpha,\delta) \equiv \avgX{\detMXad} = \Sntot^{-4} \avgX{(\TdataX)^4 \left(D^X(\alpha,\delta)\right)^2}
 \,.
\end{equation}
With this convention, an empirical tuning of $\Ftho$, as described in section~VI.B of
\cite{keitel2014:_linerobust}, can be expected to yield similar values as for the unweighted $\OSN$.
Numerical tests indeed show only small changes in the optimal $\Ftho$ over typical ranges in false-alarm
probabilities.

\clearpage

\subsection{Synthetic tests of the sensitivity-weighted statistics}
\label{sec:sens-weight-difftests}

In this section, we present results from synthetic-draw comparisons of the sensitivity-weighted detection
statistics $\OSLdetM$ and $\OSNdetM$ against their unweighted counterparts $\OSL$ and $\OSN$, covering a
similar range of noise populations as in \sref{sec:safety}.

\subsubsection{Gaussian noise\\}
\label{sec:sens-weight-difftests-gauss}

To determine the effect of sensitivity-weighting on the detection performance of the line-robust statistics,
we first revisit the same case as covered in \fref{fig:safety-rocs-H1H2-gauss}, but including
additional sensitivity ratios.
For a colocated network of H1 and H2, signals with $\snrS=4$ and pure Gaussian noise, the corresponding set of
ROCs is shown in \fref{fig:sens-weight-improvement-rocs-H1H2-gauss}.

Panel (a) of \fref{fig:sens-weight-improvement-rocs-H1H2-gauss} shows the case of equal sensitivity, where
$\OSLdetM$ and \mbox{$\OSNdetMten$} perform exactly as their unweighted counterparts.
This is expected from the analytical expressions in \eref{eq:sens-weight-OSLdetM} and
\eref{eq:sens-weight-OSNdetM}, as in this case $\sweightX=1$ and so the statistics revert back to the
unweighted forms, \eref{eq:OSL} and \eref{eq:OSN}.

For increasing ratios of \mbox{$\sqrtSnLHOtwo/\sqrtSnLHO \in \{2,5,10\}$} (panels (b)--(d) of
\fref{fig:sens-weight-improvement-rocs-H1H2-gauss}),
there is still no difference between $\OSN$ and $\OSNdetM$ with both at $\Ftho=10$.
However, $\pDet$ for the unweighted $\OSL$ decreases, while $\OSLdetM$ actually improves and approaches the
performance of $2\F$ and \mbox{$\OSNten$}.
ROCs for intermediate values of $\Ftho$ would fall between the curves for $\OSL$ (which corresponds to
\mbox{$\Ftho\rightarrow-\infty$}) and \mbox{$\Ftho=10$}.

\subsubsection{Lines in the less sensitive detector\\}
\label{sec:sens-weight-difftests-H2lines}

The case of an H1-H2 network with \mbox{$\sqrtSnLHOtwo=10\sqrtSnLHO$} and a line contamination of
\mbox{$\linefrac^{\LHOtwo}=0.1$} in the weaker detector was considered before in
\sref{sec:safety-H2lines} for the unweighted statistics, and the results were similar to pure
Gaussian noise:
significant losses in $\pDet$ for $\OSL$ and $\OSN$ for low $\Ftho$, while \mbox{$\OSNten$} in this case is
still completely safe when compared to the $\F$-statistic.

\begin{figure}[t!]
 \newsavebox{\FigBoxDetProbDiffsHHVarLineSnrWeakLines}
 \sbox{\FigBoxDetProbDiffsHHVarLineSnrWeakLines}{\includegraphics[width=0.5\textwidth]{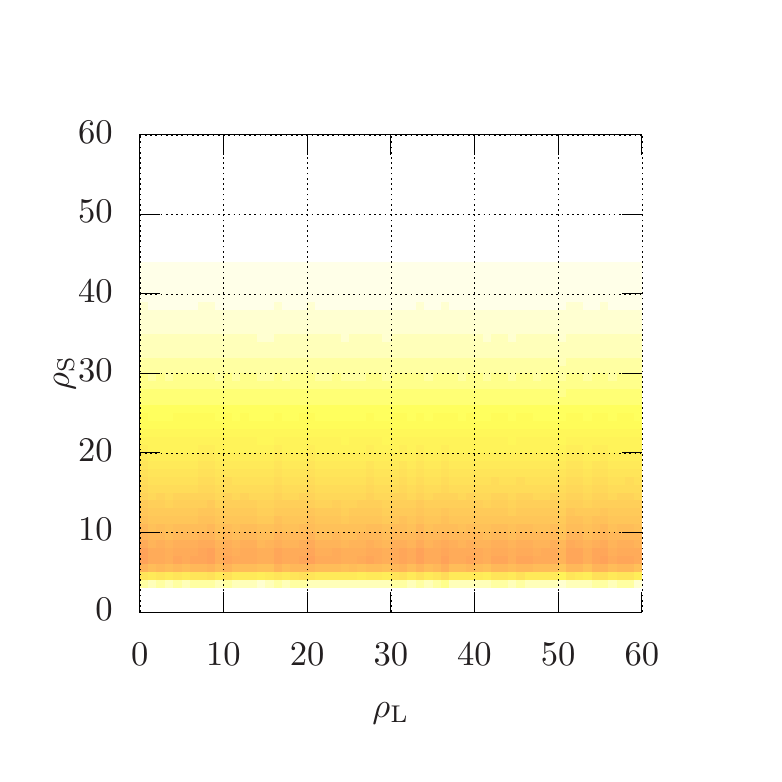}}
 \begin{minipage}[b]{0.5\textwidth}
  \raggedright (a)\\\vspace*{-1.25cm}
  \usebox{\FigBoxDetProbDiffsHHVarLineSnrWeakLines}
 \end{minipage}
 \hspace{-0.5cm}
 \begin{minipage}[b]{0.05\textwidth}
  \includegraphics[height=0.833\ht\FigBoxDetProbDiffsHHVarLineSnrWeakLines]{synth_detprob_diffs_colorbar} 
  \vspace{-2.9cm} 
 \end{minipage}
 \hspace{0.5cm}
 \begin{minipage}[b]{0.5\textwidth}
  \raggedright (b)\\\vspace*{-1.25cm}
  \includegraphics[width=\textwidth]{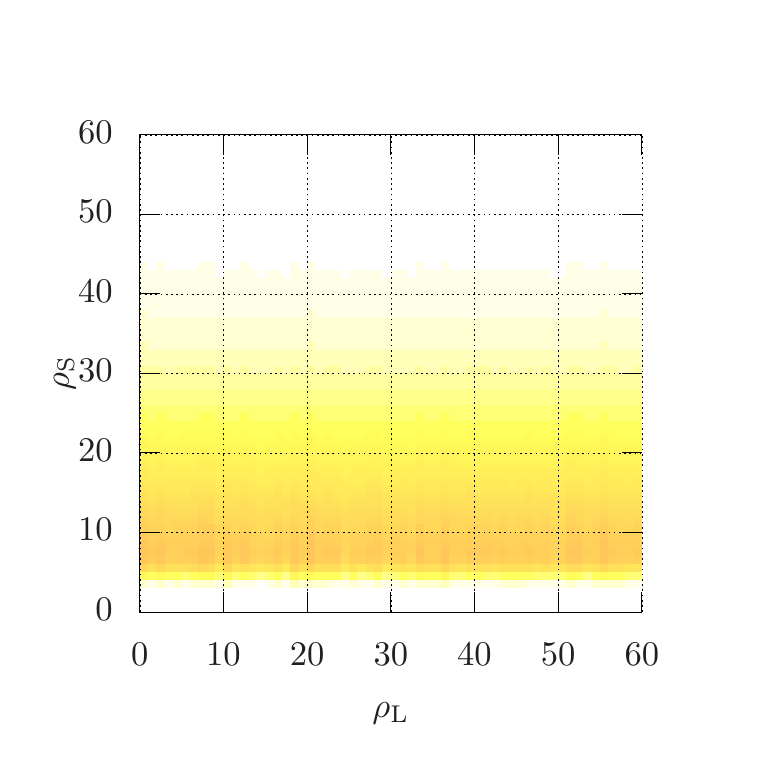}
 \end{minipage}\\[-1cm]
 \begin{minipage}[b]{0.5\textwidth}
  \raggedright (c)\\\vspace*{-0.75cm} 
  \includegraphics[width=\textwidth]{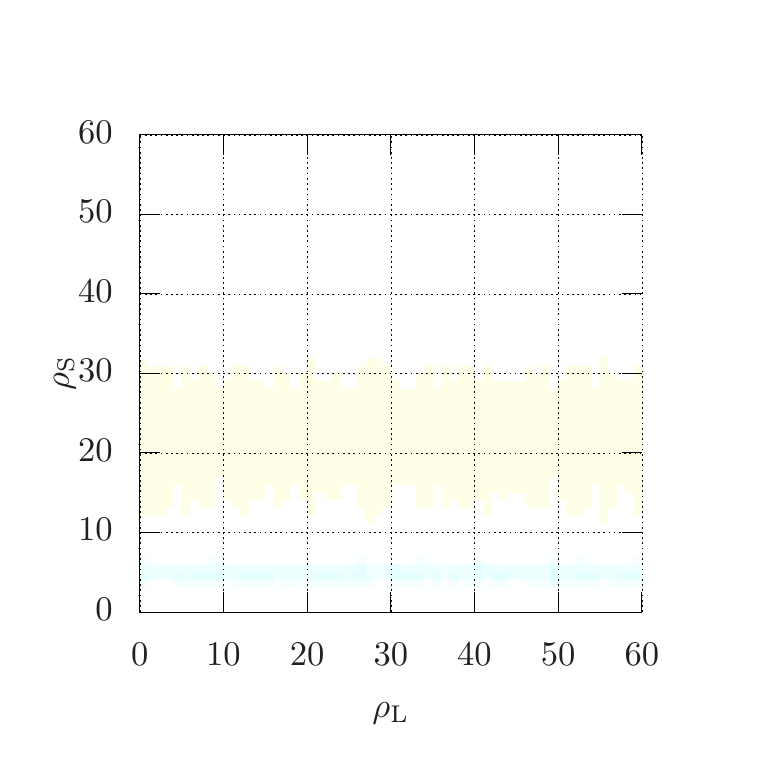}
 \end{minipage}
 \hspace{-0.5cm}
 \begin{minipage}[b]{0.05\textwidth}
  \mbox{} 
 \end{minipage}
 \hspace{0.5cm}
 \begin{minipage}[b]{0.5\textwidth}
  \raggedright (d)\\\vspace*{-0.75cm}
  \includegraphics[width=\textwidth]{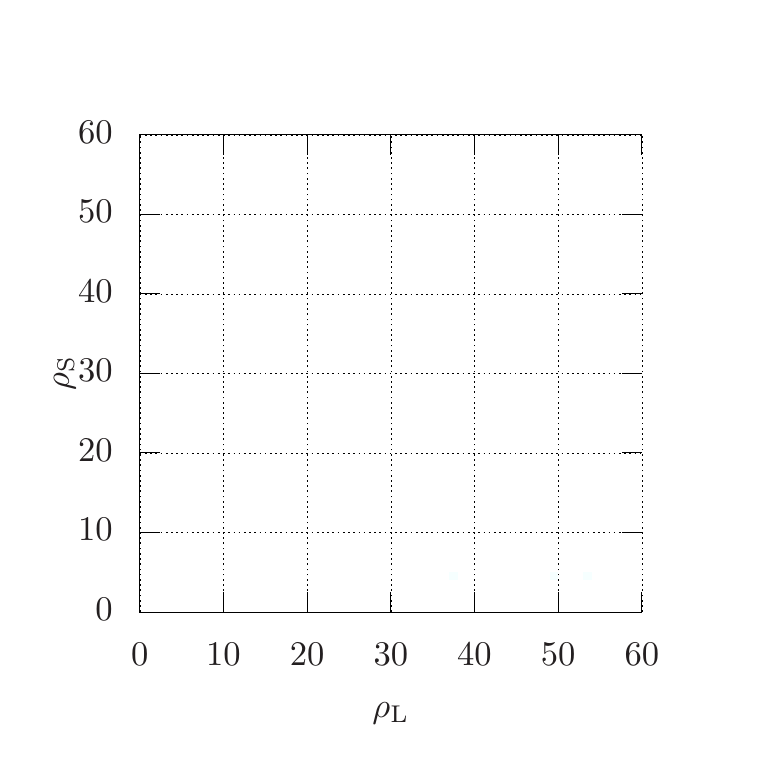}
 \end{minipage}
 \vspace{-\baselineskip}
 \caption{
  \label{fig:sens-weight-improvement-linesH2-paramspace-varlinesnr-pFA0.001-detprobdiffs}
  Lines in the less sensitive detector:
  Detection-probability differences between weighted and unweighted statistics,
  for an H1-H2 network with \mbox{$\sqrtSnLHOtwo=10\sqrtSnLHO$},
  as a function of line SNR $\snrL$ and signal SNR $\snrS$,
  at fixed \mbox{$\linefrac^{\LHOtwo}=0.1$} and \mbox{$\pFA=0.001$}. \newline
  \mbox{(a): $\pDet(\OSLdetM)-\pDet(\OSL)$},
  \mbox{(b): $\pDet(\OSNdetM)-\pDet(\OSN)$} at \mbox{$\Ftho=0$}, \newline
  \mbox{(c): $\pDet(\OSNdetM)-\pDet(\OSN)$} at \mbox{$\Ftho=4$},
  \mbox{(d): $\pDet(\OSNdetM)-\pDet(\OSN)$} at \mbox{$\Ftho=10$}.
 }
\end{figure}

To illustrate the effect of sensitivity weighting for this noise population,
\fref{fig:sens-weight-improvement-linesH2-paramspace-varlinesnr-pFA0.001-detprobdiffs} shows differences of
$\pDet$ between weighted and unweighted statistics over the same range in $\snrS$ and $\snrL$ as in
\fref{fig:safety-H1H2-linesH2-paramspace-varlinesnr-pFA0.001-detprobs}.
Similarly to the Gaussian-noise ROCs in \fref{fig:sens-weight-improvement-rocs-H1H2-gauss}, $\OSLdetM$ and
\mbox{$\OSNdetMnull$} can regain 20-30\% of $\pDet$ in comparison to their unweighted counterparts, whereas
for high $\Ftho$ such as 10 the changes are negligible.

\subsubsection{Lines in the more sensitive detector\\}
\label{sec:sens-weight-difftests-H1lines}

On the other hand, as we have already seen in \fref{fig:safety-rocs-H1H2-linesH1}, the detection
problem is more difficult in general if there are lines present in the more sensitive detector.
These can be very hard to distinguish from CW signals with only the information in $\{\F,\F^X\}$.

\begin{figure}[h!tbp]
 \newsavebox{\FigBoxDetProbDiffsHHVarLineSnrStrongLines}
 \sbox{\FigBoxDetProbDiffsHHVarLineSnrStrongLines}{\includegraphics[width=0.5\textwidth]{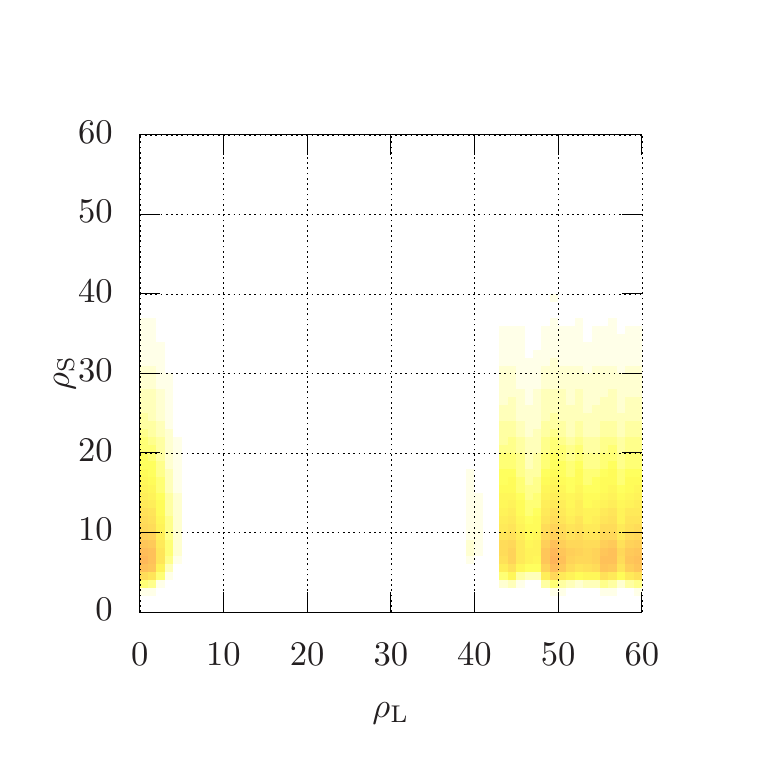}}
 \begin{minipage}[b]{0.5\textwidth}
  \raggedright (a)\\\vspace*{-1.25cm}
  \usebox{\FigBoxDetProbDiffsHHVarLineSnrStrongLines}
 \end{minipage}
 \hspace{-0.5cm}
 \begin{minipage}[b]{0.05\textwidth}
 \includegraphics[height=0.833\ht\FigBoxDetProbDiffsHHVarLineSnrStrongLines]{synth_detprob_diffs_colorbar} 
  \vspace{-2.9cm} 
 \end{minipage}
 \hspace{0.5cm}
 \begin{minipage}[b]{0.5\textwidth}
  \raggedright (b)\\\vspace*{-1.25cm}
  \includegraphics[width=\textwidth]{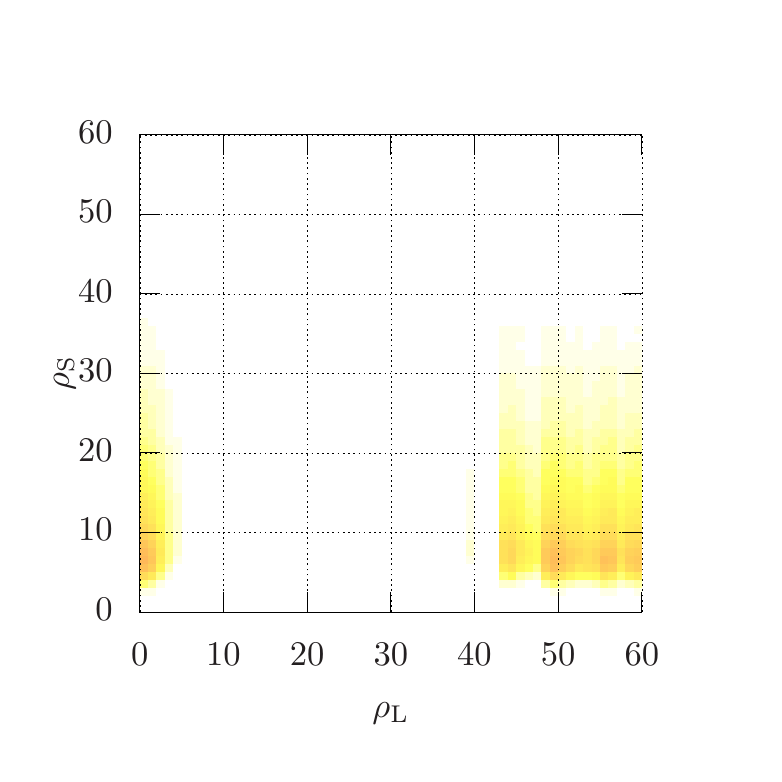}
 \end{minipage}\\[-1cm]
 \begin{minipage}[b]{0.5\textwidth}
  \raggedright (c)\\\vspace*{-0.75cm}
  \includegraphics[width=\textwidth]{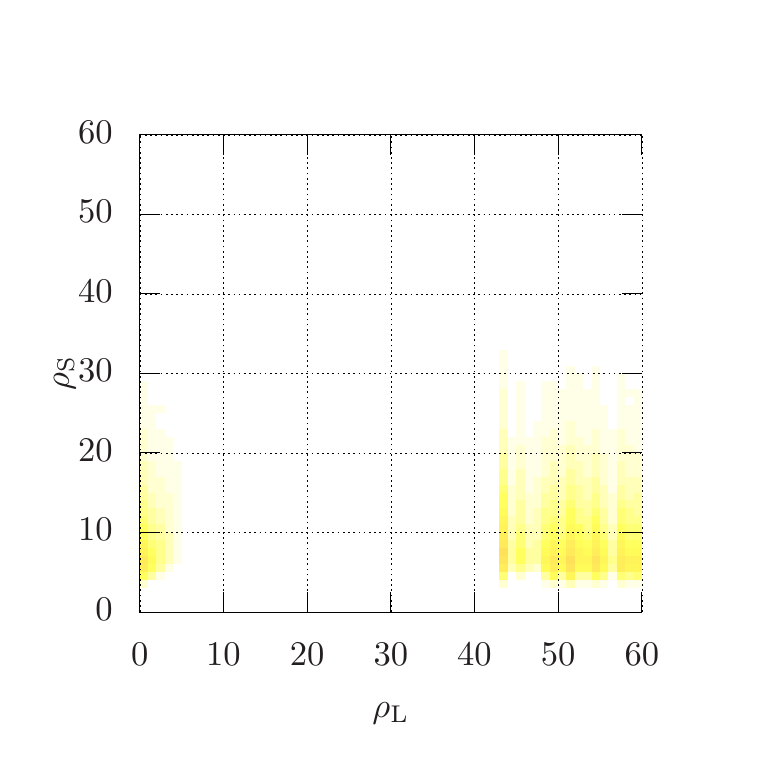}
 \end{minipage}
 \hspace{-0.5cm}
 \begin{minipage}[b]{0.05\textwidth}
  \mbox{} 
 \end{minipage}
 \hspace{0.5cm}
 \begin{minipage}[b]{0.5\textwidth}
  \raggedright (d)\\\vspace*{-0.75cm}
  \includegraphics[width=\textwidth]{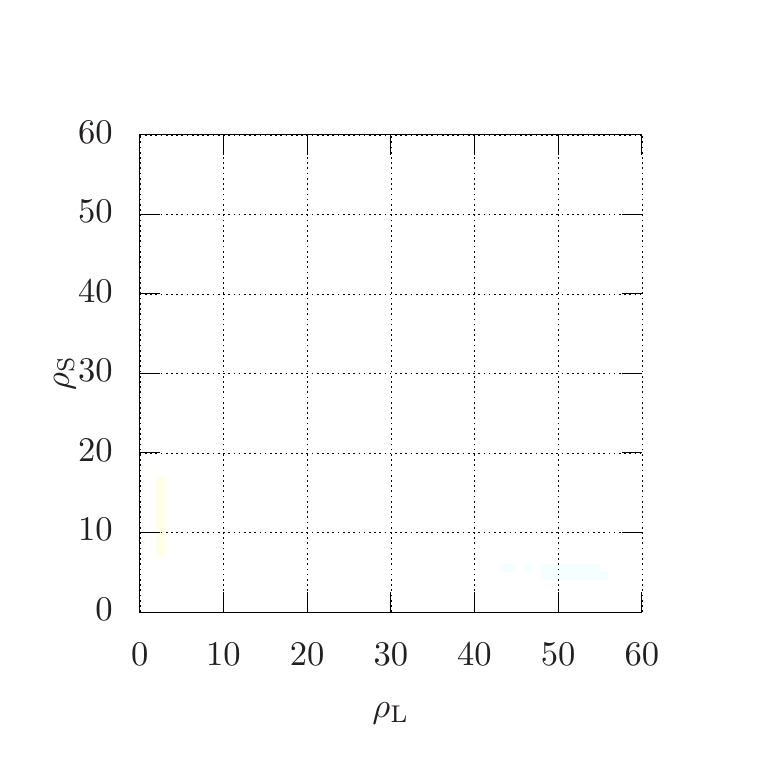}
 \end{minipage}
 \vspace{-\baselineskip}
 \caption{
  \label{fig:sens-weight-improvement-linesH1-paramspace-varlinesnr-pFA0.001-detprobdiffs}
  Lines in the more sensitive detector:
  Detection probability differences between weighted and unweighted statistics,
  for an H1-H2 network with \mbox{$\sqrtSnLHOtwo=10\sqrtSnLHO$},
  as a function of line SNR $\snrL$ and signal SNR $\snrS$,
  at fixed \mbox{$\linefrac^{\LHO}=0.1$} and \mbox{$\pFA=0.001$}. \newline
  \mbox{(a): $\pDet(\OSLdetM)-\pDet(\OSL)$},
  \mbox{(b): $\pDet(\OSNdetM)-\pDet(\OSN)$} at \mbox{$\Ftho=0$}, \newline
  \mbox{(c): $\pDet(\OSNdetM)-\pDet(\OSN)$} at \mbox{$\Ftho=4$},
  \mbox{(d): $\pDet(\OSNdetM)-\pDet(\OSN)$} at \mbox{$\Ftho=10$}.
 }
\end{figure}

For the same parameters as in
\fref{fig:safety-H1H2-linesH1-paramspace-varlinesnr-pFA0.001-detprobs}, i.e. an H1-H2 network with
\mbox{$\sqrtSnLHOtwo=10\sqrtSnLHO$} and a line contamination of \mbox{$\linefrac^{\LHO}=0.1$}, changes in
$\pDet$ due to sensitivity-weighting are shown in
\fref{fig:sens-weight-improvement-linesH1-paramspace-varlinesnr-pFA0.001-detprobdiffs}.
In these noise populations, the improvements of the sensitivity-weighted counterparts to $\OSL$ and $\OSN$
with low $\Ftho$ are more modest than in the previous cases, with $\pDet$ improving only in the low- and
high-$\snrL$ regions, and not in the most problematic range in-between.
Sensitivity weighting again brings no improvement for an optimally-tuned \mbox{$\OSNten$}.

\subsubsection{Sky-location dependence\\}
\label{sec:sens-weight-difftests-H1L1}

We have also compared synthetic ROCs for the weighted and unweighted statistics for a non-colocated network of
H1 and L1 at the 'worst-case' different-sensitivity sky locations discussed in \sref{sec:safety-H1L1}.

For $\Tobs=12\,\hours$, the differences are already negligible in pure Gaussian noise and small, at most a few
percent, in the presence of lines in either detector.
All differences are completely negligible for $\Tobs=24\,\hours$.

\section{Conclusions}
\label{sec:sens-weight-conclusions}

The synthetic tests presented in this paper lead to the following main conclusions about the safety of
line-robust statistics for unequally sensitive detectors, and about the simple sensitivity-weighting attempt
to improving it:
\begin{enumerate}
 \item The sky-location-dependent differences in detector sensitivities, due to different antenna patterns,
       are generally too small to lead to any noticeable effects.
       Even for a search over a short observation time of 12\,h and directed at a location with extreme
       differences in antenna patterns, the effect on detection probabilities is small.
       Hence, the line-robust detection statistics from \cite{keitel2014:_linerobust} can be considered as
       just as safe for directed searches as for the all-sky searches tested before, provided the detectors
       have similar noise PSDs and amounts of data.
 \item To notice any effects due to unequal sensitivities, these must be quite pronounced, for example a ratio
       in $\sqrtSnX$ of well over 2, which is larger than typical values encountered for the LIGO H1-L1
       network.
 \item Even for very unequal sensitivities (for example a factor 10 in $\sqrtSnX$), the line-robust statistic
       $\OSN$ of \eref{eq:OSN} with an $\Ftho$-tuning as described in section~VI.B of
       \cite{keitel2014:_linerobust} is still \emph{safe} in Gaussian noise -- in the sense of not being worse
       than the $\F$-statistic -- and improves over $\F$ in the presence of strong lines in a less sensitive
       detector only.
 \item The line-veto statistic $\OSL$ of \eref{eq:OSL}, as well as $\OSN$ with lower
       values of $\Ftho$, are \emph{not safe} in these cases.
       Here, sensitivity-weighting can recover some losses.
       However, there would normally be no reason to use these statistics in place of the tuned $\OSN$, for
       which sensitivity-weighting makes no significant difference in any of the cases considered here.
 \item For very unequal sensitivities, lines in the most sensitive detector lead to partial losses and partial
       gains of the tuned $\OSN$ compared to the $\F$-statistic.
       The cases where $\OSN$ is unsafe (at high signal SNRs) are arguably of less practical relevance than
       those were it still yields an improvement (for weaker signals and very strong lines).
       Again, sensitivity-weighting makes no difference to the performance of the tuned $\OSN$.
\end{enumerate}
Thus, we find that that the sensitivity-weighting approach through the use of a modified amplitude prior, as
described in \sref{sec:sens-weight}, does not seem a promising direction in practice.
The ``unweighted'' but tuned line-robust statistic, as described in \cite{keitel2014:_linerobust}, is
generally found to be safe even for detectors with highly unequal sensitivities.
The remaining weakness for particularly ``line-like signals'' might be a subject for further work, although
the quantitative limits demonstrated here make this seem less urgent for practical applications.

\section*{Acknowledgements}
We thank the following colleagues for insightful discussions and comments: Bruce Allen, Berit Behnke,
Heinz-Bernd Eggenstein, Maria Alessandra Papa, Keith Riles, Karl Wette, John T. Whelan and Graham Woan.
DK was supported by the IMPRS on Gravitational Wave Astronomy.
This paper has been assigned LIGO document number \dcc{} and AEI-preprint number \aei{}.


\appendix
\section{Details on antenna patterns}
\label{sec:antpat-appendix}

For reference, here we give the full expressions for the antenna-pattern matrix and functions, following the
notation of \cite{prix:_cfsv2}.

Consider a GW propagating in the direction of the normal vector $-\uvn$ in a reference frame fixed in the
solar system barycentre (SSB), with an axis $\uvz$ orthogonal to the equatorial plane.
Then, the propagating wave frame is given by unit vectors
\begin{equation}
 -\uvn \,, \quad
 \uvxi = \uvn \cross \uvz / \left| \uvn \cross \uvz \right|
 \quad \mathrm{and} \quad
 \uveta = \uvxi \cross \uvn
\end{equation}
and polarisation tensors
\begin{eqnarray}
 \epsp &= \uvxi \otimes \uvxi - \uveta \otimes \uveta \\
 \epsc &= \uvxi \otimes \uveta + \uveta \otimes \uvxi \,.
\end{eqnarray}
For a detector $X$ whose position on the earth's surface is given, again in SSB coordinates, by the tensor
$d^X$, the two antenna pattern functions corresponding to the two polarisation components are then given, in
the long-wavelength limit, by
\begin{eqnarray}
 \label{eq:antpats}
  a^X(t,\uvn) &\equiv d^X_{ij}(t) \, \epsp^{ij}(\uvn) \,, \\
  b^X(t,\uvn) &\equiv d^X_{ij}(t) \, \epsc^{ij}(\uvn) \,.
\end{eqnarray}
The antenna-pattern matrix components $A$, $B$, $C$ of \eref{eq:Mmunu} are defined as \emph{noise-weighted}
averages over \emph{Short Fourier Transforms} (SFTs):
\begin{eqnarray}
 A &= \avgSFTs{a^2(t)} \nonumber \,, \\
 B &= \avgSFTs{b^2(t)} \,, \\
 C &= \avgSFTs{a(t)\,b(t)} \nonumber \,.
\end{eqnarray}
The averages are performed according to
\begin{equation}
 \label{eq:SFTavg_noiseweighted}
 \avgSFTs{Q} = \frac{1}{\Nsft} \sum\limits_{X=1}^{\Ndet} \sum\limits_{\alpha=1}^{\Nsft} \nwXal \, Q_{X\alpha}
 \,,
\end{equation}
with per-SFT and per-detector \emph{noise-weights} factors
\begin{equation}
 \label{eq:noiseweights}
 \nwXal = \frac{\Sntot}{\SnXal}
\end{equation}
which fulfil the normalisation constraint
\begin{equation}
 \sum_{X\alpha} \nwXal = \sum_{X}\NsftX = \Nsft \,.
\end{equation}
Notably, for single-detector quantities $Q^X$, we could use noise-weights normalised by $\SnX$ instead of
$\Sntot$; however, the convention of \cite{prix:_cfsv2} and current LALSuite implementations
\footnote{\url{https://www.lsc-group.phys.uwm.edu/daswg/projects/lalsuite.html}}
is to use the same averaging as for multi-detector quantities:
\begin{equation}
 \label{eq:SFTavg_noiseweighted_X}
 \avgSFTs{Q^X} = \frac{1}{\NsftX} \sum\limits_{\alpha=1}^{\NsftX} \nwXal \, Q^X_{\alpha} \,,
\end{equation}
so that
\begin{eqnarray}
 A^X &= \frac{1}{\NsftX} \sum\limits_{\alpha=1}^{\NsftX} \nwXal \, a^{X2}(t) \nonumber \\
 B^X &= \frac{1}{\NsftX} \sum\limits_{\alpha=1}^{\NsftX} \nwXal \, b^{X2}(t)  \\
 C^X &= \frac{1}{\NsftX} \sum\limits_{\alpha=1}^{\NsftX} \nwXal \, a^X(t)\,b^X(t) \nonumber \,.
\end{eqnarray}
This normalisation means that the per-detector antenna-pattern matrix determinants are again given by
multiplying with the \emph{multi}-detector sensitivity and data volume:
\begin{equation}
 \label{eq:detMX-appendix}
 \detMX = \left(\SinvT\right)^4 D^{X2} \,.
\end{equation}
\begin{figure}[h!tbp]
 \newsavebox{\FigBoxAntPatsHalfDay}
 \sbox{\FigBoxAntPatsHalfDay}{\includegraphics[width=0.5\textwidth]{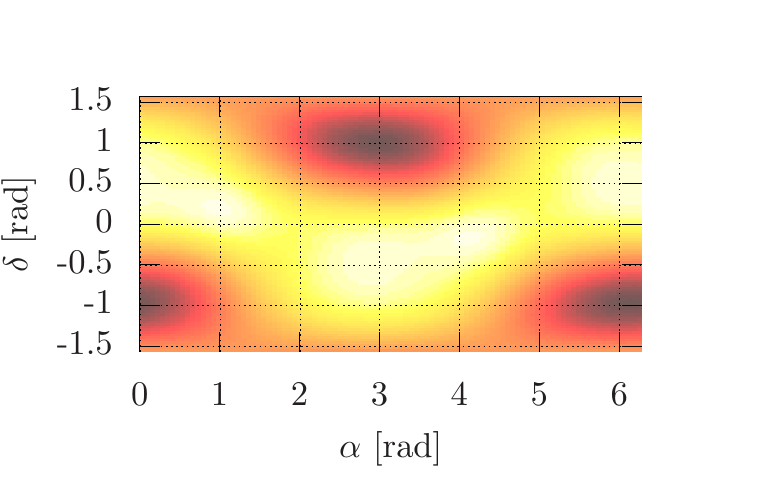}}
 \begin{minipage}[b]{0.5\textwidth}
  \raggedright H1\\\vspace*{-1.0cm}
  \usebox{\FigBoxAntPatsHalfDay}
 \end{minipage}
 \hspace{-0.5cm}
 \begin{minipage}[b]{0.05\textwidth}
 \includegraphics[height=1.53\ht\FigBoxAntPatsHalfDay]{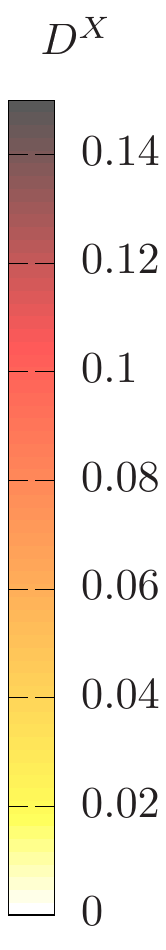} 
  \vspace{-2.9cm} 
 \end{minipage}
 \hspace{0.5cm}
 \begin{minipage}[b]{0.5\textwidth}
  \raggedright L1\\\vspace*{-1.0cm}
  \includegraphics[width=\textwidth]{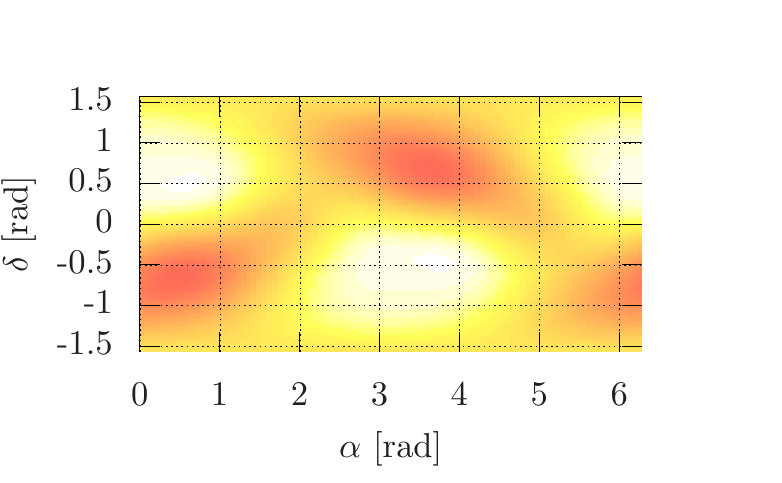}
 \end{minipage}\\[-0.25cm]
 \begin{minipage}[b]{0.5\textwidth}
  \raggedright G1\\\vspace*{-1.0cm}
  \includegraphics[width=\textwidth]{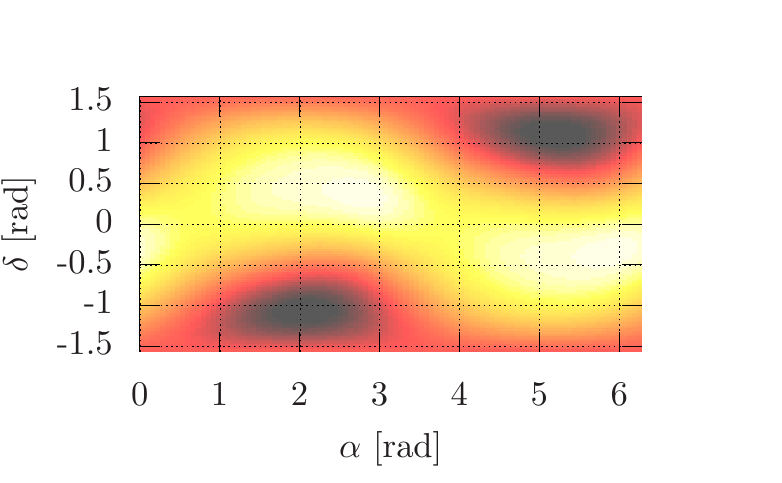}
 \end{minipage}
 \hspace{-0.5cm}
 \begin{minipage}[b]{0.05\textwidth}
  \mbox{} 
 \end{minipage}
 \hspace{0.5cm}
 \begin{minipage}[b]{0.5\textwidth}
  \raggedright V1\\\vspace*{-1.0cm}
  \includegraphics[width=\textwidth]{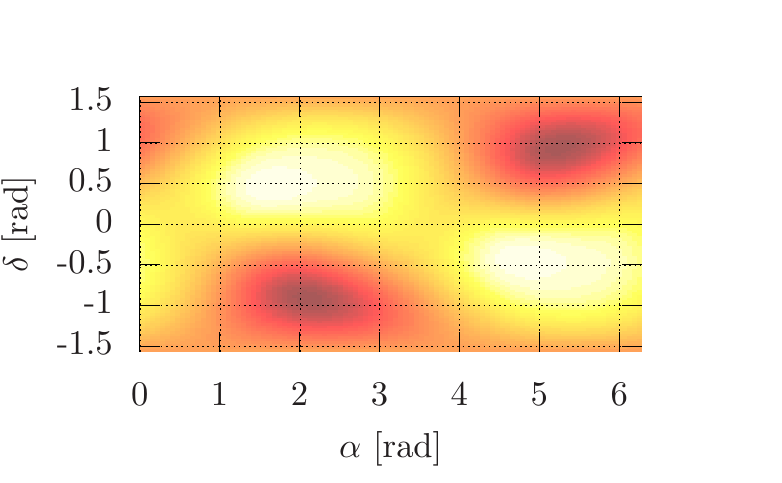}
 \end{minipage}
 \caption{
  \label{fig:antpatD_allsky_12h}
  Per-detector antenna pattern matrix determinant $D^X$ (colour scale)
  over the sky ($\alpha$, $\delta$ in rectangular projection)
  averaged over 12 hours (24 SFTs)
  starting from GPS time 852443819 (Jan 10, 2007, 05:56:45 UTC; during the LIGO S5 run)
  for
  LIGO Hanford (H1),
  LIGO Livingston (L1),
  GEO600 (Hannover, G1),
  Virgo (Cascina, V1).
 }
\vspace{4\baselineskip}
 \newsavebox{\FigBoxAntPatsFullDay}
 \sbox{\FigBoxAntPatsFullDay}{\includegraphics[width=0.5\textwidth]{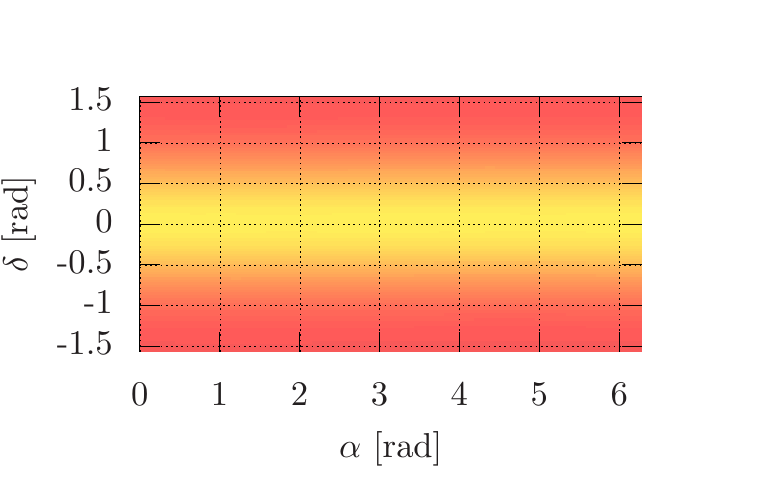}}
 \begin{minipage}[b]{0.5\textwidth}
  \raggedright H1\\\vspace*{-1.0cm}
  \usebox{\FigBoxAntPatsFullDay}
 \end{minipage}
 \hspace{-0.5cm}
 \begin{minipage}[b]{0.05\textwidth}
 \includegraphics[height=1.53\ht\FigBoxAntPatsFullDay]{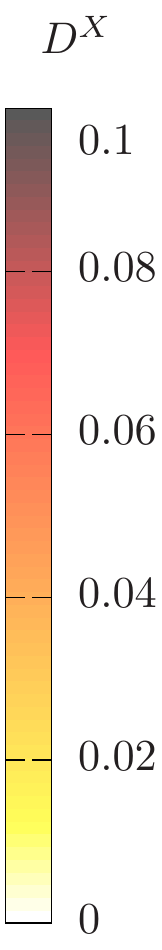} 
  \vspace{-2.9cm} 
 \end{minipage}
 \hspace{0.5cm}
 \begin{minipage}[b]{0.5\textwidth}
  \raggedright L1\\\vspace*{-1.0cm}
  \includegraphics[width=\textwidth]{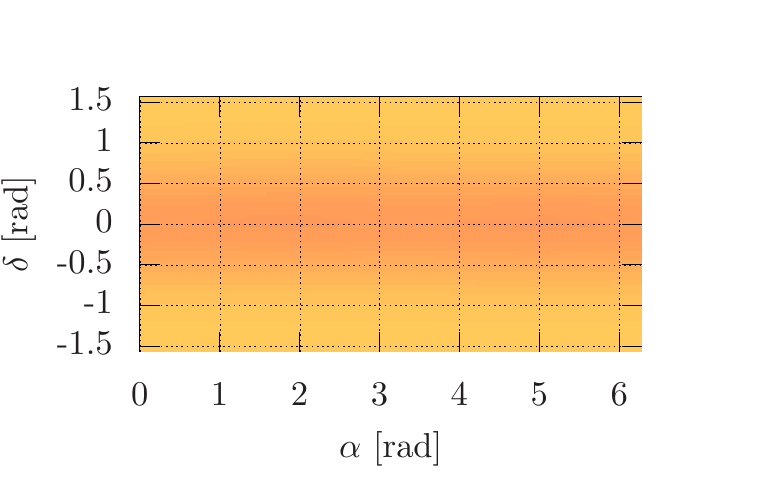}
 \end{minipage}\\[-0.25cm]
 \begin{minipage}[b]{0.5\textwidth}
  \raggedright G1\\\vspace*{-1.0cm}
  \includegraphics[width=\textwidth]{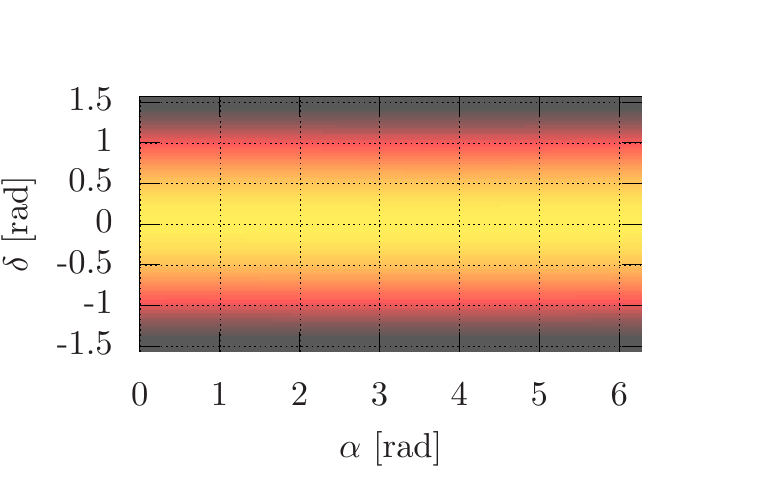}
 \end{minipage}
 \hspace{-0.5cm}
 \begin{minipage}[b]{0.05\textwidth}
  \mbox{} 
 \end{minipage}
 \hspace{0.5cm}
 \begin{minipage}[b]{0.5\textwidth}
  \raggedright V1\\\vspace*{-1.0cm}
  \includegraphics[width=\textwidth]{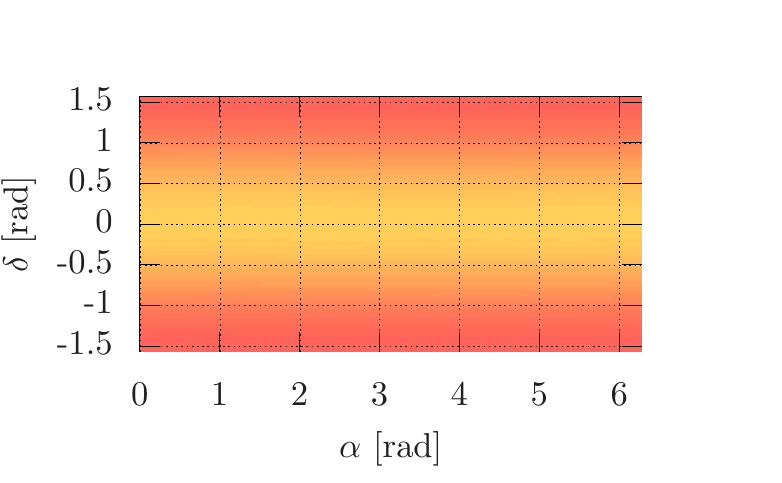}
 \end{minipage}
 \caption{
  \label{fig:antpatD_allsky_24h}
  Per-detector antenna pattern matrix determinant $D^X$ (colour scale)
  over the sky ($\alpha$, $\delta$ in rectangular projection)
  averaged over 24 hours (48 SFTs)
  starting from GPS time 852443819 (Jan 10, 2007, 05:56:45 UTC; during the LIGO S5 run)
  for
  LIGO Hanford (H1),
  LIGO Livingston (L1),
  GEO600 (Hannover, G1),
  Virgo (Cascina, V1). \newline
  Note that the small remaining $\alpha$-dependence would only average out over a \emph{sidereal} day ($\approx23.9344\,$hours) and using finer time resolution.
 }
\end{figure}
Although this quantity is invariant under a normalisation change like, for example,
\begin{equation}
 \nwXal \rightarrow \nwXal' = \frac{\SnX}{\SnXal} \,,
\end{equation}
the $A^X$, $B^X$, $C^X$ and $D^X$ are not, and the determinant transforms as
\begin{equation}
 \detMX = \left(\SinvT\right)^4 D^{X2} =  \left(\SinvTX\right)^4 D'^{X2}  \,.
\end{equation}

In \fref{fig:antpatD_allsky_12h}, we see a sky-map of $D$ for four different detectors (LIGO H1 in
Hanford, Washington, USA; LIGO L1 in Livingston, Louisiana, USA; GEO600 near Hannover, Germany and Virgo in
Cascina, Italy) averaged over 12 hours.

In contrast, \fref{fig:antpatD_allsky_24h} adds another 12 hours.
We see that for a whole day of observation (or integer multiples thereof), most of the variation in right
ascension $\alpha$ is averaged away.

{
\small
\bibliography{../../biblio.bib}
}

\end{document}

%% file: defs2.tex
\def\mbf#1{\ensuremath{\mathchoice{\mbox{\boldmath$\displaystyle#1$}}
{\mbox{\boldmath$\textstyle#1$}}
{\mbox{\boldmath$\scriptstyle#1$}}
{\mbox{\boldmath$\scriptscriptstyle#1$}}}}

\newcommand{\hours}{\mathrm{h}}

\newcommand{\prob}[2]{P\left(#1|#2\right)}
\newcommand{\probI}[1]{P\left(#1\right)}
\newcommand{\scalar}[2]{\left(#1|#2\right)}

\newcommand{\detVec}[1]{\mbf{#1}}

\newcommand{\prior}[1]{\MakeLowercase{#1}}

\newcommand{\avgX}[1]{\left\langle #1 \right\rangle_X}	
\newcommand{\avgSFTs}[1]{\left\langle #1 \right\rangle_{\mathrm{SFTs}}}	

\newcommand{\Dop}{\lambda}

\newcommand{\Amp}{\mathcal{A}}

\newcommand{\hmax}{h_{\mathrm{max}}}
\newcommand{\rhomax}{\widehat{\rho}_{\mathrm{max}}}
\newcommand{\rhomaxS}{\widehat{\rho}_{\mathrm{max}\Signal}}
\newcommand{\rhomaxL}{\widehat{\rho}_{\mathrm{max}\Line}}
\newcommand{\rhomaxLX}{\widehat{\rho}_{\mathrm{max}\Line}^X}
\newcommand{\cF}{c_*}

\newcommand{\M}{\mathcal{M}}
\newcommand{\MX}{\mathcal{M}^X}

\newcommand{\detM}{\left|\M\right|}
\newcommand{\detMX}{\left|\MX\right|}

\newcommand{\Sn}{S}	
\newcommand{\SnX}{\Sn^X}		
\newcommand{\SnXal}{\Sn^X_\alpha}	
\newcommand{\Sntot}{\mathcal{S}}
\newcommand{\sqrtSnX}{\sqrt{\Sn^X}}		

\newcommand{\dVx}{\detVec{x}}
\newcommand{\dVy}{\detVec{y}}
\newcommand{\dVn}{\detVec{n}}
\newcommand{\dVh}{\detVec{h}}

\newcommand{\Hyp}{\mathcal{H}}
\newcommand{\Gauss}{\mathrm{\MakeUppercase{G}}}
\newcommand{\Signal}{{\mathrm{\MakeUppercase{S}}}}
\newcommand{\Line}{{\mathrm{\MakeUppercase{L}}}}
\newcommand{\Noise}{{\Gauss\Line}}
\newcommand{\HypS}{\Hyp_\Signal}
\newcommand{\HypN}{\Hyp_\Noise}
\newcommand{\HypL}{\Hyp_\Line}
\newcommand{\HypG}{\Hyp_\Gauss}

\providecommand{\sc}[1]{\widehat{#1}}
\renewcommand{\sc}[1]{\widehat{#1}}

\newcommand{\OSN}{O_{\Signal\Noise}}	
\newcommand{\OSG}{O_{\Signal\Gauss}}	
\newcommand{\OSL}{O_{\Signal\Line}}

\newcommand{\OLG}{O_{\Line\Gauss}}
\newcommand{\oLGX}{o_{\Line\Gauss}^X}

\newcommand{\oSG}{\prior{\OSG}}
\newcommand{\oSL}{\prior{\OSL}}
\newcommand{\oLG}{\prior{\OLG}}
\newcommand{\oSN}{\prior{\OSN}}

\newcommand{\lineprob}{p_\Line}
\newcommand{\linefrac}{f_\Line}

\newcommand{\F}{\mathcal{F}}		

\newcommand{\Ftho}{\F_*^{(0)}}
\newcommand{\FX}{\F^X}
\newcommand{\rX}{r^X}

\newcommand{\pDet}{p_{\mathrm{det}}}	
\newcommand{\pFA}{p_{\mathrm{FA}}}

\newcommand{\OR}{\;\mathrm{or}\;}

\newcommand{\Ndet}{{N_{\mathrm{det}}}}

\newcommand{\const}{\mathrm{const}}

\newcommand{\eto}{\ensuremath{\mathrm{e}^}}	

\newcommand{\sft}{{\mathrm{SFT}}}

\newcommand{\Tdata}{T_{\mathrm{data}}}
\newcommand{\TdataX}{T_{\mathrm{data}}^X}

\newcommand{\Nsft}{N_\sft}
\newcommand{\NsftX}{N_\sft^X}

\newcommand{\LHO}{\mathrm{H1}}
\newcommand{\LHOtwo}{\mathrm{H2}}
\newcommand{\LLO}{\mathrm{L1}}

\newcommand{\snr}{\rho} 
\newcommand{\snrS}{\snr_{\Signal}}
\newcommand{\snrL}{\snr_{\Line}}

\newcommand{\Sinv}{\Sntot^{-1}}
\newcommand{\SinvT}{\Sinv \Tdata}

\newcommand{\SinvTf}{\Sntot^{-4} \Tdata^4}
\newcommand{\STinvsq}{\Sntot^{2} \Tdata^{-2}}
\newcommand{\SinvX}{\left(\SnX\right)^{-1}}
\newcommand{\SinvTX}{\frac{\TdataX}{\SnX}}

\newcommand{\cross}{\times}
\newcommand{\epsp}{\epsilon_{+}}
\newcommand{\epsc}{\epsilon_{\cross}}
\newcommand{\unitvect}[1]{\widehat{#1}}
\newcommand{\uvn}{\unitvect{n}}
\newcommand{\uvz}{\unitvect{z}}
\newcommand{\uvxi}{\unitvect{\xi}}
\newcommand{\uveta}{\unitvect{\eta}}

\newcommand{\SdetM}{\Signal^{\M}}
\newcommand{\LdetM}{\Line^{\M}}
\newcommand{\NdetM}{{\Gauss\LdetM}}
\newcommand{\HypSdetM}{\Hyp_{\SdetM}}
\newcommand{\HypLdetM}{\Hyp_{\LdetM}}
\newcommand{\HypNdetM}{\Hyp_{\NdetM}}

\newcommand{\OSdetMG}{O_{\SdetM\Gauss}}

\newcommand{\OSLdetM}{O_{\Signal\LdetM}}

\newcommand{\OSNdetM}{O_{\Signal\NdetM}}

\newcommand{\oSdetMG}{o_{\SdetM\Gauss}}
\newcommand{\oSLdetM}{o_{\Signal\LdetM}}

\newcommand{\oSNdetM}{o_{\Signal\NdetM}}

\newcommand{\nw}{w}

\newcommand{\nwXal}{\nw_{X\alpha}}

\newcommand{\avD}{\mathcal{D}}

\newcommand{\bcoh}{$\widetilde{\textrm{b}}$}
\newcommand{\sqrtSnLHO}{\sqrt{\Sn^{\LHO}}}
\newcommand{\sqrtSnLHOtwo}{\sqrt{\Sn^{\LHOtwo}}}
\newcommand{\sqrtSnLLO}{\sqrt{\Sn^{\LLO}}}
\newcommand{\Ndraws}{{N_{\mathrm{draws}}}}
\newcommand{\OSNnull}{\OSN(\Ftho=0)}
\newcommand{\OSNten}{\OSN(\Ftho=10)}
\newcommand{\OSNdetMnull}{\OSNdetM(\Ftho=0)}
\newcommand{\OSNdetMten}{\OSNdetM(\Ftho=10)}
\newcommand{\sweightX}{q^X}
\newcommand{\detMXad}{\left|\MX(\alpha,\delta)\right|}
\newcommand{\Tobs}{T_{\mathrm{obs}}}

\newcommand{\detA}{\mathrm{A}}
\newcommand{\detB}{\mathrm{B}}
